\documentclass[journal,twoside,web]{ieeecolor}
\usepackage{generic}
\usepackage{cite}
\usepackage{amsmath,amssymb,amsfonts}
\usepackage{algorithmic}
\usepackage{graphicx}
\usepackage{textcomp}
\usepackage{xprintlen}
\usepackage{url}
\usepackage{hyperref}       % hyperlinks % [hidelinks]
\hypersetup{
    colorlinks=true,
    citecolor=blue,
    linkcolor=red,
    filecolor=magenta,      
    urlcolor=violet}
\usepackage{xcolor, soul}
\sethlcolor{pink}
\usepackage{array}

\usepackage{xparse, mathtools, array}
\DeclarePairedDelimiterX{\rvect}[1]{[}{]}{\,\makervect{#1}\,}
\ExplSyntaxOn
\NewDocumentCommand{\makervect}{m}
 {
  \seq_set_split:Nnn \l_tmpa_seq { , } { #1 }
  \begin{matrix}
  \seq_use:Nn \l_tmpa_seq { & }
  \end{matrix}
 }
\ExplSyntaxOff
\newcommand{\Transp}{\mathsf{T}}

\newcommand{\beginsupplement}{%
        \setcounter{table}{0}
        \renewcommand{\thetable}{S-\Roman{table}}%
        \setcounter{figure}{0}
        \renewcommand{\thefigure}{S-\arabic{figure}}%
     }
\usepackage{bbm}
\usepackage{multirow}
\usepackage{tabularx}
\usepackage{cleveref}
\usepackage{csvsimple}
\usepackage{numprint}
\npdecimalsign{.}
\usepackage{layouts}
\usepackage{float}
\usepackage{caption}
\usepackage{booktabs}

\Crefname{figure}{Fig.}{Figs.}
\crefname{figure}{Fig.}{Figs.}
\def\BibTeX{{\rm B\kern-.05em{\sc i\kern-.025em b}\kern-.08em
    T\kern-.1667em\lower.7ex\hbox{E}\kern-.125emX}}
    
\begin{document}
% reference shortening to et al.
% \bstctlcite{IEEEexample:BSTcontrol}

\title{ECG-Based Electrolyte Prediction: Evaluating Regression and Probabilistic Methods}
\author{Philipp Von Bachmann, Daniel Gedon, Fredrik K. Gustafsson, Ant\^{o}nio H. Ribeiro, Erik Lampa, \\Stefan Gustafsson, Johan Sundström, and Thomas B. Schön 
\thanks{\emph{This work has been submitted to the IEEE for possible publication. Copyright may be transferred without notice, after which this version may no longer be accessible.}}
\thanks{The study was funded by The Kjell and Märta Beijer Foundation; Anders Wiklöf; the Wallenberg AI, Autonomous Systems and Software Program (WASP) funded by Knut and Alice Wallenberg Foundation; Uppsala University; and has received funding from the European Research Council (ERC) under the European Union's Horizon Europe research and innovation programme (grant agreement n° 101054643). The computations were enabled by resources in project sens2020005 and sens2020598 provided by the Swedish National Infrastructure for Computing (SNIC) at UPPMAX, partially funded by the Swedish Research Council through grant agreement no. 2018-05973.}
\thanks{Philipp Von Bachmann is with the Dept. of Computer Science, University of Tübingen, Germany.}
\thanks{Daniel Gedon, Fredrik K. Gustafsson, Ant\^{o}nio H. Riberio and Thomas B. Schön are with the Dept. of Information Technology, Uppsala University, Sweden (e-mail \{daniel.gedon,fredrik.gustafsson\}@it.uu.se).}
\thanks{Erik Lampa, Stefan Gustafsson and Johan Sundström are with the Clinical Epidemiology Unit, Dept. of Medical Sciences, Uppsala University, Sweden. Johan Sundström is also with the George Institute for Global Health, University of New South Wales, Sydney, Australia.}
}

\maketitle

\begin{abstract}
    \emph{Objective:} Imbalances of the electrolyte concentration levels in the body can lead to catastrophic consequences, but accurate and accessible measurements could improve patient outcomes. While blood tests provide accurate measurements, they are invasive and the laboratory analysis can be slow or inaccessible. In contrast, an electrocardiogram (ECG) is a widely adopted tool which is quick and simple to acquire. However, the problem of estimating continuous electrolyte concentrations directly from ECGs is not well-studied. We therefore investigate if regression methods can be used for accurate ECG-based prediction of electrolyte concentrations.
\emph{Methods:} We explore the use of deep neural networks (DNNs) for this task. We analyze the regression performance across four electrolytes, utilizing a novel dataset containing over 290\thinspace000 ECGs. For improved understanding, we also study the full spectrum from continuous predictions to binary classification of extreme concentration levels. To enhance clinical usefulness, we finally extend to a probabilistic regression approach and evaluate different uncertainty estimates.
\emph{Results:} We find that the performance varies significantly between different electrolytes, which is clinically justified in the interplay of electrolytes and their manifestation in the ECG. We also compare the regression accuracy with that of traditional machine learning models, demonstrating superior performance of DNNs.
\emph{Conclusion:} Discretization can lead to good classification performance, but does not help solve the original problem of predicting continuous concentration levels. While probabilistic regression demonstrates potential practical usefulness, the uncertainty estimates are not particularly well-calibrated.
\emph{Significance:} Our study is a first step towards accurate and reliable ECG-based prediction of electrolyte concentration levels.

\end{abstract}


\section{Introduction}
\label{section:introduction}
Electrolytes such as potassium or calcium influence the water and acid-base balance in the human body and ensure the proper functioning of muscles, brain and heart \cite{EDELMAN1959256,CARLSON2008529}. Imbalances in the electrolyte concentrations are common among hospitalized patients. For example, approximately 25\,\% have been found to have abnormal potassium levels \cite{paice1986record,el2011electrolyte}. Such imbalances can lead to serious heart conditions, ranging from arrhythmia to cardiac arrest \cite{fisch1973relation}.

Electrolyte imbalances are often only detected by analyzing blood tests, since symptoms rarely appear until the imbalance is at a severe level. Blood tests provide accurate measurements of the electrolyte concentration levels, but are invasive and the laboratory analysis can be slow or not even accessible in remote locations. Since electrolytes directly influence the functioning of the heart, there are known but complex relationships between some electrolytes and the electrocardiogram (ECG) \cite{SURAWICZ1967814}. An ECG measures the electrical activity of the heart, it is low-cost and a widely available routine diagnostic tool for heart-related conditions in both primary and specialized care. If accurate measurements of electrolyte concentration levels could be automatically extracted directly from an ECG, it would give access to non-invasive, convenient and quick electrolyte monitoring for individuals in a large population.

Computer-based, automatic processing of ECGs is a well established technology~\cite{macfarlane2005university}. Recently, as a promising alternative to traditional methods which utilize hand-crafted features in combination with simple models, deep neural networks (DNNs) achieved strong performance on classification of cardiac diseases with known ECG patterns \cite{rajpurkar2017cardiologist,hannun2019cardiologist, ribeiro2020automatic}. Interestingly, DNNs also demonstrated promising results on detecting patterns that are not easily identifiable by traditional electrocardiographic analysis. For instance, models that detect myocardial infarction in ECG exams without ST-elevation~\cite{gustafsson_nstemi}, and recent work on accurate models capable of predicting the risk of mortality~\cite{raghunath2020prediction, lima2021deep}, atrial fibrillation~\cite{attia_artificial_2019,biton_atrial_2021} and left ventricular dysfunction~\cite{attia_screening_2019} directly from the ECG. Most of these models are predictive of the outcome even for seemingly normal ECGs.

ECG-based classification problems using DNNs are thus extensively studied, and DNNs have been demonstrated to outperform traditional machine learning models in this setting \cite{biton_atrial_2021,kwon2021artificial}. We are therefore interested in applying DNNs to the problem of automatic electrolyte prediction. Electrolyte concentration levels are however \emph{continuous} values. Thus, electrolyte prediction is naturally formulated as a \emph{regression} problem. There are many regression methods using DNNs in the general literature \cite{gast2018lightweight, varamesh2020mixture, xiao2018simple, ruiz2018fine, gustafsson2020energy}, but few have been applied to the problem of ECG-based electrolyte prediction. This includes the most common and straightforward approach of \emph{deep direct regression} \cite{lathuiliere2019comprehensive}, in which a DNN directly outputs continuous predictions and is trained by minimizing the mean-squared error between predicted and observed values. Previous work instead either only used hand-crafted ECG features together with simple models~\cite{frohnert1970statistical, velagapudi2017computer}, performed the simplified task of classifying abnormal hypo (low) and hyper (high) concentration levels \cite{galloway2019development,kwon2021artificial}, or discretized the concentration levels and used a method similar to ordinal regression \cite{lin2020deep}. Moreover, Lin et al.~\cite{lin2020deep} only studied prediction of a single electrolyte (potassium).

In this work, we therefore study in detail how DNNs can be used to predict the continuous concentration levels of electrolytes directly from ECGs. We start by applying the deep direct regression approach, and analyze its regression accuracy across four important electrolytes. We here find that the performance varies significantly between different electrolytes. We also compare the performance with that of traditional models such as Gradient Boosting and Random Forest, demonstrating superior performance of DNNs also in our setting of ECG-based regression. To conduct this analysis, we utilize a novel large-scale dataset of over $290\thinspace000$ ECGs.

In cases when deep direct regression fails to accurately predict the continuous level, we study the prediction problem in more detail. We discretize the electrolyte concentration level and train classification models, with an increasing number of classes, studying the full spectrum: from continuous prediction to binary classification of extreme concentration levels. This provides insights on the inherent difficulty of the prediction problem, and enables us to extract as fine-grained predictions as possible, for different electrolytes. When we instead observe that the direct regression model learns a clear relationship between inputs and targets, we also attempt to improve the clinical usefulness of the regression model by extending to a \emph{probabilistic regression} approach \cite{kendall2017uncertainties}, providing uncertainty estimates for the predictions. We evaluate these uncertainty estimates on both in-distribution and out-of-distribution data.

Our contributions can be summarized as:
\begin{itemize}
    \item We utilize a novel large-scale dataset of over $290\thinspace000$ ECGs from adult patients attending emergency departments at Swedish hospitals.
    
    \item We train deep direct regression models for ECG-based prediction of continuous electrolyte concentration levels, demonstrating that DNNs outperform traditional machine learning models on this task.

    \item We explore probabilistic regression approaches, providing the continuous predictions with uncertainty estimates.
    
    \item We analyze the regression accuracy across four important electrolytes: potassium, calcium, sodium and creatinine\footnote{While potassium, calcium and sodium are by definition electrolytes, creatinine is an abundant blood biomarker. For the ease of reading on descriptions, we do however denote creatinine as an electrolyte in this study.}, finding that the performance varies significantly.
    
    \item We discretize the electrolyte concentration levels and train classification models, studying the full spectrum from continuous prediction to binary classification of extreme levels.
\end{itemize}

\section{Background}
    \label{section:Background}
    % In this paper we work with regression and and its probabilistic version. Furthermore, we perform classification by binning and by ordinal regression. The different terms and notation will be introduced first.
    
    We formulate ECG-based electrolyte prediction as a regression problem and explore different regression approaches. We also study the prediction problem further by discretizing the concentration levels and training models on the simplified classification task.% Here, we introduce the different terms and notation.

    \subsection{Regression and Uncertainty Estimation}
    \label{section:Background_Regression}
        The goal in a regression problem is to predict a continuous target variable $y\in\mathbb{R}$ for any input variable $x$, given a training dataset $\mathcal{D} = \{x_i, y_i\}_{i=1}^n$ of $n$ data points. This is achieved by training a model $m_\theta$ with parameters $\theta$ such that a loss function is minimized on $\mathcal{D}$. In the common deep direct regression approach, the model is a DNN outputting continuous predictions, $\hat{y} = m_\theta(x)$, using the mean-squared error (MSE) as loss function. These predictions do not capture any measure of uncertainty. In medical applications, where predictions might impact the treatment of patients, this lack of uncertainty is especially problematic.
    
        Probabilistic regression aims to solve this problem by estimating different types of uncertainties \cite{gal2016thesis, kendall2017uncertainties, lakshminarayanan2017simple}. (1) \textit{Aleatoric uncertainty} captures irreducible ambiguity from the experiment itself. One example is noise from a measurement device. (2) \textit{Epistemic uncertainty} refers to a lack of knowledge and is therefore reducible. Out-of-distribution (OOD) data is one example where epistemic uncertainty is expected to be high.
        
        Aleatoric uncertainty can be estimated by explicitly modelling the conditional distribution $p(y|x)$. Assuming a Gaussian likelihood leads to the parametric model $p(y \vert x; \theta)=\mathcal{N}\big( y; \mu_\theta(x) , \sigma^2_\theta(x)\big)$, where the DNN $m_\theta$ outputs both the mean $\mu$ and variance $\sigma^2$, i.e.\, $m_\theta(x) = \rvect{\mu_{\theta}(x), \sigma^2_{\theta}(x)}^{\Transp}$. % \in \mathbb{R}^2$.
        The mean is used as a target prediction, $\hat{y} = \mu_\theta(x)$, whereas the variance $\sigma^2_{\theta}(x)$ is interpreted as an estimate of input-dependent aleatoric uncertainty. The DNN is trained by minimizing the corresponding negative log-likelihood $- \sum_{i=1}^{n} \log p(y_i | x_i; \theta)$.
        
        This approach does however not capture epistemic uncertainty. One way to add this uncertainty is by treating the model parameters $\theta$ according to the Bayesian framework \cite{neal1995bayesian} and learning a \emph{posterior} probability distribution over the parameters. Ensemble methods \cite{dietterich2000ensemble, lakshminarayanan2017simple} constitute a simple approach to estimate epistemic uncertainty. Multiple models are trained and the uncertainty is estimated by the variance of the prediction over all models. Ensemble methods usually improve the regression accuracy of the model and have been shown to be highly competitive baselines for uncertainty estimation methods \cite{ovadia2019can,gustafsson2020evaluating}.
        Another common method is the Laplace approximation \cite{mackay1992bayesian}, which approximates the \emph{posterior} distribution $p(\theta \vert \mathcal{D})$ with a Gaussian,
        \begin{equation} \notag
            p(\theta \vert \mathcal{D}) \approx \mathcal{N}(\theta ; \theta_{\text{MAP}}, \Sigma),
        \end{equation}
        where the inverse of the covariance matrix $\Sigma^{-1}=-\nabla_{\theta}^2 \log p(\mathcal{D}, \theta) \vert_{\theta_\text{MAP}}$ is the negative Hessian matrix, evaluated at the \emph{maximum-a-posteriori} estimate $\theta_{\text{MAP}}$. Laplace approximations can be applied post-hoc to a pretrained model with reduced computational complexity \cite{kristiadi2020being, daxberger2021laplace}.

    \subsection{Simplifying Regression via Discretization}
    \label{Background:RegressionThroughClassification}
        If accurate prediction of the continuous target variables $y~\in~\mathbb{R}$ is unachievable or even not required, a regression problem can be simplified into a standard classification problem by discretizing the targets. Specifically, the target range needs to be divided into $k$ intervals and each target~$y$ assigned to the respective interval \cite{torgo1996regression}. The model now outputs a distribution over $k$ classes and predictions are made by the class with maximum probability.
  
        Usually, the Cross-Entropy (CE) loss is used to train the model, which can however lead to rank inconsistency of the original continuous problem. This implies that for a predicted class~$\tilde{y}$ corresponding to a certain target interval, the probability for the classes corresponding to neighbouring intervals do not necessarily decrease monotonically away from the predicted class. To address this issue, Cao et al.~\cite{cao2020rank} proposed rank-consistency \emph{ordinal regression}. The output of the model is changed to denote the probability that the target $y$ is larger than or equal to the lower bound of the corresponding interval of class $j$. The model is trained with binary CE loss, and class predictions are computed as $\tilde{y} = 1 + \sum_{j=1}^{k} p_\theta(x)_j$, where $p_\theta(x)_j = \mathbbm{1}_{m_\theta(x)_j > 0.5}$.

    \subsection{Clinical Relevance}
        In an in-hospital scenario, blood measurement with laboratory analysis is the gold standard to accurately and reliably determine the electrolyte concentration levels of patients. However, there are multiple scenarios where ECG-based prediction models are desired.
        
        First, in the ambulance setting, there is typically no access to onboard blood laboratory analysis equipment but it is possible to acquire ECGs. Many countries apply a telehealth setup and send the ECGs to a coronary care unit for reading and decision making. If the patient in the ambulance has presented with arrhythmia, which is potentially lethal, it is important to quickly identify its cause. If electrolyte disturbances could be estimated, and identified as the cause of the arrhythmia, then life-saving treatment could be started directly in the ambulance. For example, an insulin-glucose infusion, an intravenous calcium injection, or an inhalation of a beta-2-agonist are all treatments for hyperkalemia (high potassium) that could be administered. Onboard automated ECG analysis would be highly useful in this scenario.
        
        Second, in rural areas without specialists a remote setting with telehealth care centers is built up. One such example is the Telehealth Network of Minas Gerais, Brazil \cite{Alkmim2012} which receives up to 5\thinspace000~ECGs per day. In many locations in that network, obtaining an ECG may be easier than obtaining a blood sample for electrolyte analysis. Our model could provide crucial care for these patients which would otherwise not be possible at all.
        
        Third, in an in-hospital setting, an ECG-based electrolyte prediction could be useful for monitoring treatment of hyperkalemia or hypernatremia (high sodium). For hyperkalemia, monitoring that potassium is decreased fast enough is an objective; for hypernatremia, monitoring that sodium is decreased slowly enough (to prevent potentially lethal brain edema due to osmosis) is an important objective. A real-time ECG-based prediction could replace frequent blood draws in these scenarios.

        Extending the prediction model with uncertainty estimation increases its clinical usefulness. Above we defined two types of uncertainty, each of which plays a crucial role. (1)~\textit{Aleatoric uncertainty} captures inherent ambiguity in the data itself. For example due to measurement noise, it is inherently more difficult to determine the electrolyte levels for some ECGs than for others. Hence, an accurate prediction of concentration level might not be possible for some ECGs. A model that properly captures aleatoric uncertainty could automatically detect such cases, enabling doctors to take appropriate action such as acquiring a new ECG or asking for blood test analysis instead. If a model captures (2)~\textit{epistemic uncertainty}, it could detect cases when the ECG being analysed during clinical deployment is out-of-distribution compared to the training data. Failing to detect such cases could lead to highly incorrect model predictions, with potentially catastrophic consequences, since the accuracy of DNN models can drop significantly on out-of-distribution examples \cite{hendrycks2018benchmarking, koh2021wilds}.
\section{Related Work}

Most previous work on ECG-based electrolyte prediction rely on hand-crafted ECG features. These are specific characteristics such as the time between two waves, amplitude or slope of a wave. Frohnert et al.~\cite{frohnert1970statistical} were the first to manually develop a relation between such features and electrolyte concentrations. More recently, \cite{attia2016novel,corsi2017noninvasive,velagapudi2017computer} rely on hand-crafted features but automatically fit the model parameters to data. Their model performance is however still limited. DNNs offer a different approach by instead learning both features and predictions jointly. Convolutional neural networks (CNNs) have shown promising results for classifying different ECG patterns \cite{rajpurkar2017cardiologist,hannun2019cardiologist,ribeiro2020automatic}. 
For electrolyte prediction, Galloway et al.~\cite{galloway2019development} used an 11-layer CNN to classify hyperkalemia. Lin et al.~\cite{lin2020deep} were the first to develop a DNN for regression on potassium, using an approach similar to ordinal regression by discretizing the model outputs. Hence, despite recent work on ECG-based predictions for electrolytes, it remains unclear if the common deep direct regression approach can be applied to accurately predict electrolyte concentration levels from ECGs. 

%\cite{lin2020deep} were the first to develop a DNN for regression on potassium, using an approach similar to ordinal regression by discretizing the model outputs. Hence, despite recent work on ECG-based predictions for electrolytes, it remains unclear if the common deep direct regression approach can be applied to accurately predict electrolyte concentration levels from ECGs. 

Little work has gone into deep prediction models for electrolytes other than potassium. Kwon et al.~\cite{kwon2021artificial} studied potassium, calcium and sodium, but they only considered the simplified problem of classifying hypo and hyper conditions. We are the first to study these electrolytes in the original problem setting of regressing continuous concentration levels. Moreover, we also consider prediction of creatinine. We further apply probabilistic regression methods for uncertainty estimation. Closest to this probabilistic setting is \cite{xia2021benchmarking}, which proposed dataset shifts and compared the change of uncertainty for different models, but concentrated exclusively on classification problems.
\section{Methods}
\label{section:Methods}
%\Antonio{Maybe add subsections to methods: Dataset, Model,...}

%\Antonio{I found this next paragraph out of place: this does not sounds like methods, but like contribution. I would remove or move it}
%In this work, we address the lack of regression for electrolyte prediction. Additionally, we train ordinal regression and interval based classification models to show their performance when regression is too difficult \hl{(clarify TODO!)}.

% We extract a new large-scale dataset that links ECGs to four different electrolytes. Then, we train individual models for each electrolyte with different prediction methods.

We extract a new large-scale dataset that links ECGs to four different important electrolytes. Then, we train regression and classification models for each electrolyte. The study has been approved by relevant ethical review authorities.

\subsection{Dataset}
\label{section:dataset}

%%%%%%%%%%%%%%%%%%%%%%%%%%%%%%%%%%%%%%%%%%%%%%%%%%%%%%5
% Here, describe the dataset and some statistics and set in relation to datasets from other works out of related work section \Daniel{add figures of dataset stats in appendix and refer to it (e.g. ECG collection vs time, age/sex distribution, \dots}

% preprocessing
% \begin{itemize}
%     \item duration of ecg/number of leads (only 8 independent)
%     \item sampling rate/padding
%     \item baseline removal/power line noise removal
% \end{itemize}

% probably do a table with at least
% - count
% - mean, std deviation
% - years
% - locations (roughly)
% - sex
% - age

% maybe we want to add statistics by metadata, especially if we include later in model
% here maybe a plot for:
% electrolyte by age, maybe even subdivided by sex 

% \Fredrik{Also compile dataset stats from the other papers (how many examples did they train on etc.), compare with these, e.g. in a table. Put this somewhere, perhaps in the appendix.}

% what data do we use?
We use data from adult patients attending six emergency departments in the Stockholm area, Sweden, between 2009 and 2017. The ECG recordings are linked through unique patient identifiers to blood measurements of electrolyte concentration levels of potassium, calcium, sodium and creatinine, extracted from electronic health records with laboratory measurements. % from the Stockholm region.
Inclusion filters are applied to only include data where the ECG and blood measurement are acquired within $\pm60~$minutes. Larger time frames would enable more patients in our study, but at the cost of lower label quality.

% how do we pre-process the data
Standard 10-seconds 12-lead ECGs are recorded, where we use the 8 independent leads since the remaining ones are mathematically redundant. The data is sampled, producing an ECG trace of size $channels \times samples$. We pre-process all ECG recordings to a sampling frequency of $400~$Hz and pad with zeros to obtain $4\thinspace096$ samples. We further apply a high-pass filter to remove biases and low-frequency trends, and finally remove possible power line noise using a notch filter. 
The ground truth electrolyte concentration levels are obtained by blood tests. Details on pre-processing are provided in \Cref{appendix:dataset prepro}.

% data splitting
%\Antonio{I think there is some mistake in this paragraph. If I remember correct the split is regarding the patients, not the ecgs.}
We split our datasets into training, validation and test sets. $70\,\%$ of the patients are used for model development including training and validation. The remaining $30\,\%$ are split into $20\,\%$ for a \textit{random test set}, where the recorded ECGs overlap in time with the development set. The last $10\,\%$ are used for a \textit{temporal test set}, where the ECGs do not overlap in time with the development set, which is used to observe changes in recordings over time. We removed patients from the temporal test set who already had recordings in other datasets to avoid data leakage. In the main paper we present results on the random test set with the exception of \Cref{table:Regression}, while complementing results for the temporal test set are in the appendix. %The temporal test set is our best external test set, since we do not have access to other ECG and electrolyte concentration data. \Daniel{Maybe remove last sentence and replace by "In the main paper we present results on the random test set, while the results for the temporal test set can be found in Appendix XYZ." But we still have to disuss it in the discussion.}

% statistics of the data
We obtain four datasets -- one for each electrolyte. The number of patients range between $79\thinspace577$ and $166\thinspace908$, with between $126\thinspace970$ and $295\thinspace606$ ECGs in total, see \Cref{table:dataset} for more characteristics. Sometimes multiple blood measurements and multiple ECGs are recorded within the selected $\pm60$~minute time frame. We select the median electrolyte value and assign it to all ECGs for training.  We consider multiple ECGs during training as a form of data augmentation. In the validation and test sets, we use only the first ECG. Details and comparisons with datasets from literature are in \Cref{appendix:dataset char}.

\setlength{\tabcolsep}{2pt}
\begin{table}[t]
    % \normalsize
    \centering
    \caption{Characteristics of our datasets.}
    \begin{tabular}{ll|rrrr}
        \toprule
         && Potassium & Calcium & Sodium & Creatinine \\
        \midrule
        Patients && 165\thinspace508 & 79\thinspace577 & 163\thinspace610 & 166\thinspace908 \\[2pt]
        Recordings && 290\thinspace889 & 125\thinspace970 & 288\thinspace891 & 295\thinspace606 \\[2pt]
        \% Male& & 49.38 & 48.71 & 49.07 & 49.22 \\[2pt]
        \multirow{2}{40pt}{Age}
        & mean & 61.26 & 60.47 & 61.41 & 61.34 \\
        & sd & 19.61 & 20.03 & 19.69 & 19.61\\[2pt]
        \multirow{2}{40pt}{Minutes diff (abs)}
        & mean & 16.28 & 12.68 & 15.92 & 16.24 \\
        & sd & 15.04 & 14.05 & 14.91 & 15.01 \\[2pt]
        \multirow{2}{45pt}{Concentration}
        & mean & 3.99 & 2.29 & 138.93 & 90.55\\
        & sd & 0.50 & 0.13 & 3.82 & 71.00\\
        \bottomrule
    \end{tabular}
    \label{table:dataset}\vspace{-1.5mm}
\end{table}
% reset
\setlength{\tabcolsep}{6pt}

\subsection{Models and Training Procedures}
\label{section:methods:model_training}

We compare the performance of DNN models with three traditional machine learning models: linear regression, Gradient Boosting \cite{friedman2001greedy} and Random Forest \cite{breiman2001random}. The raw ECG trace of size $channels \times samples$ is the input to all models. In order to keep the three traditional models computationally tractable, the ECG trace is flattened into a vector and dimensionality-reduced using principal component analysis (PCA). Based on the PCA eigenvalue distribution (see \cref{fig:Cum_Ex_Var}), we set the reduced dimensionality to 256. For the DNN models, such pre-processing is not necessary.

For the choice of DNN architecture, the literature reviews in \cite{Ebrahimi_2020,Xiong_2022} note that convolutional models such as ResNets are the dominant deep architecture for ECG-based prediction modelling. The authors in \cite{ribeiro2020automatic} also experimented with vectogram linear transformation for dimensionality reduction, LSTMs and VGG convolutional architecture, but ended up using a ResNet. Hence, we use the ResNet backbone network from \cite{ribeiro2020automatic,lima2021deep} as a feature extractor in all DNN models. Our methodological approach is however model-agnostic and any architecture with high performance could be utilized instead.

For deep direct regression, the DNN $m_\theta$ consists of the ResNet backbone and a network head that outputs target predictions, $\hat{y} = m_\theta(x)$. The DNN is trained from scratch using the MSE loss. During training we normalize the targets $y$ with z-transformation to obtain a similar target distribution across all electrolytes. We then discretize the targets into $k$ intervals, and train both classification and ordinal regression models, as described in \Cref{Background:RegressionThroughClassification}. The network head of the direct regression DNN is modified to instead output $k$ values. For ordinal regression, we train using binary CE loss and for classification using the CE loss. All models are trained for 30 epochs and the final model is selected from the best validation loss. If not specified otherwise, we train for 5 different seeds and report mean and standard deviation (sd).

For probabilistic regression, we create a Gaussian model $\mathcal{N}\big( y; \mu_\theta(x) , \sigma^2_\theta(x)\big)$ by extending the direct regression DNN with a second network head that outputs the variance $\sigma^2_\theta(x)$. The model is trained by minimizing the corresponding negative log-likelihood. We train an ensemble of 5 Gaussian models, and then extract three different uncertainty estimates: (1) \textit{Aleatoric} uncertainty is given by the average predicted variance $\sigma^2_\theta(x)$ (denoted aleatoric Gaussian). (2) \textit{Epistemic uncertainty} is computed as the variance of the predicted mean $\mu_\theta(x)$ over the 5 ensemble members (denoted epistemic ensemble). (3) We additionally define an \textit{epistemic uncertainty} by fitting a Laplace approximation after training using the \texttt{Laplace} library \cite{daxberger2021laplace} (denoted epistemic Laplace). The approximation is fit to the last layer of the mean network head by approximating the full Hessian. We report the average epistemic Laplace uncertainty over the 5 ensemble members.

The code is implemented in PyTorch \cite{paszke2019pytorch} and models are trained on a single Nvidia A100 GPU. Further training details are in \Cref{appendix:TrainingDetails}. Our complete implementation code and the trained models are publicly available at \url{https://github.com/philippvb/ecg-electrolyte-regression}.

\section{Results}
\label{section:Results}
% We first present the results for deep direct regression. Next, we perform classification and \hl{highlight that this simplification can help the performance}. Finally, we concentrate on potassium for probabilistic regression.
We first present the results for deep direct regression. Next, we compare classification and ordinal regression in the discretized regression setting. Finally, we concentrate on potassium for probabilistic regression.

\subsection{Deep Direct Regression}
    \label{Results:Regression}
    
    \begin{figure}[t]
    % Don't change width since font size is scaled to that width
        \includegraphics[width=\columnwidth]{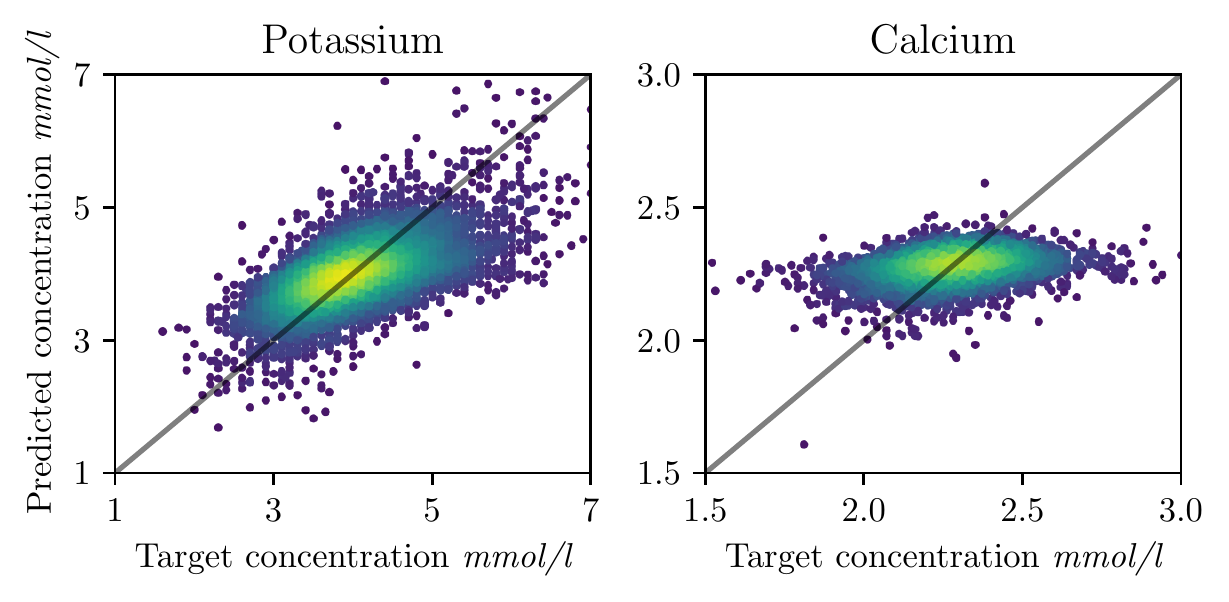}\vspace{-2.5mm} %width=0.7\textwidth (original)
        \centering
        \caption{\!\textbf{Regression scatter\! plot}:\! The diagonal depicts the optimal fit. The color indicates frequency of points (yellow: high density; blue: low density), from a Gaussian KDE.}
        \label{figure:Regression}\vspace{0.0mm}
    \end{figure}

    \setlength{\tabcolsep}{5pt}
    \begin{table}[t]
        \centering
        % \normalsize
        \caption{
        \textbf{Regression MSE/MAE}: targets are not normalized.}
        \begin{tabular}{l|ll|l}
            \toprule
            & \multicolumn{2}{c}{random test set} & temporal test set \\ 
            & MSE (sd) & MAE (sd) & MAE (sd) \\ 
            \midrule
            potassium & 0.152 (0.026) & 0.285 (0.015) & 0.262 (0.013) \\
            calcium & 0.015 ($2\mathrm{e}{-4}$) & 0.088 ($5\mathrm{e}{-4}$) & 0.1059 ($3\mathrm{e}{-4}$) \\
            %\midrule
            sodium & 12.59 (0.111) & 2.512 (0.016) & 2.390 (0.009) \\
            %\midrule
            creatinine & 3719 (86.04) & 26.69 (1.118) & 24.50 (1.298)\\
            \bottomrule 
        \end{tabular}
        \label{table:Regression}\vspace{-1.5mm}
    \end{table}
    \setlength{\tabcolsep}{6pt}

    We validate our ResNet architecture against traditional machine learning methods. For potassium, the Gradient Boosting model reaches an MSE of 0.215, the Random Forest model 0.220 and linear regression 0.220. In contrast our ResNet-based DNN reaches an MSE of 0.152, which outperforms the traditional baseline models and demonstrates benefits of using DNNs also in our setting of ECG-based regression. Hence, in the following we concentrate on the deep ResNet-based model. 
    
    Fig.~\ref{figure:Regression} depicts the results of our deep direct regression model for potassium and calcium, plotting predictions $\hat{y}$ against targets $y$. For potassium, the data points concentrate along the diagonal, indicating an overall good fit. For calcium, the predictions are horizontally aligned, meaning that the model mainly predicts the mean target value of the train dataset for all inputs $x$. %This hints that it is not able to create good predictions from the input.
    %Results for sodium and creatinine are in \cref{figure:AppendixRegressionSodiumPcreatintine} both of which are not depicting good regression fit. 
    Corresponding plots for sodium and creatinine are in \cref{figure:AppendixRegressionSodiumPcreatintine} in \Cref{appendix:additional_results}. Sodium reflects the undesirable behaviour of calcium. While creatinine shows an overall positive trend, the model seems to suffer from the high variance for higher target values, making predictions for these values noisy. Since the main behavior is captured by potassium and calcium, we will focus on them in the following. The complete results for all four electrolytes are in \Cref{appendix:additional_results}.

    The MSE and mean absolute error (MAE) in \Cref{table:Regression} do not directly reflect the performance difference between calcium and potassium, as calcium shows significantly lower errors. To understand these opposing results, we investigate the dataset distributions. The variance in the ground truth electrolyte levels is significantly lower for calcium with 0.016 compared to potassium with 0.22. Thus, predicting the mean training target value for calcium (resulting in an MSE equal to the variance) will result in lower MSE without learning the relationship between input and target. Computing errors with normalized targets gives MSE values which better reflect the performance difference, as \Cref{table:AppendixRegression} shows. 

    \begin{figure}[t]
        \centering
        \includegraphics[width=\columnwidth]{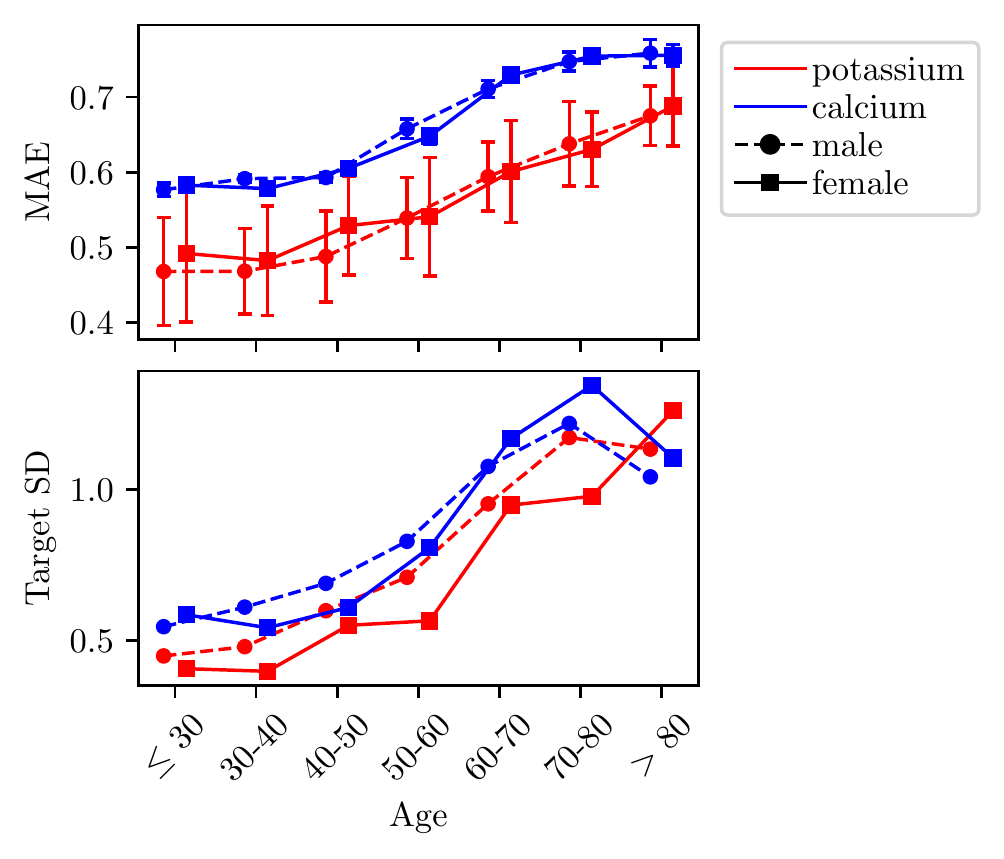}
        \caption{\textbf{Stratified regression results}: \textit{Top:} MAE of regression task stratified for different electrolytes by age and sex. \textit{Bottom:} Corresponding standard deviations of the target values stratified in the same way.}
        \label{figure:Regression_stratified}\vspace{-1.5mm}
    \end{figure}
    
    Returning to the results in \Cref{table:Regression},  we do not observe a significant drop in performance when evaluating on the temporal test set. In fact, the MAE is often slightly lower than for the random test set. This is a first indication that our model is agnostic to real-world data collection shifts. Other works that report regression performance of potassium concentrations obtain MAEs of $0.53$ (Lin et al.~\cite{lin2020deep}) and $0.50$ (Attia et al.~\cite{attia2016novel}), compared to which our model obtains superior performance. 

    % To further analyse our results, we stratify them according to age and sex in \cref{figure:Regression_stratified} (top). We observe that our results are independent of sex, but that the MAE has a positive correlation with age. Comparing with the bottom plot we see that the correlation with age is originated in a larger spread of the target values for older patients \Daniel{@Fredrik: please double check if this is sufficient for the stratification.}
    To further analyse our results, we stratify them according to age and sex in \cref{figure:Regression_stratified} (top). We observe that our results are independent of sex, but that the MAE has a positive correlation with age. Comparing with \cref{figure:Regression_stratified} (bottom) we see that this correlation is largely expected, since the variance of the ground truth target values also increases with age.

\subsection{Classification and Ordinal Regression}
%\Antonio{It is not very clear for me what is being done here. In which case is a classification model trained from scratch and in which case a threshold is used for the regression model?}
    
    % We change the direct regression to a classification task and re-train our models. By doing so, we try to simplify the predictions for electrolytes that perform poorly in direct regression. We compare both \textit{classification} and \textit{ordinal regression} for increasingly fine-grained predictions by varying the number of intervals. The class intervals are defined for each electrolyte separately: For $k$\thinspace=\thinspace3 classes we define the lower and upper interval bounds by $\mu \pm 2 \sigma$. For $k$\thinspace$>$\thinspace3 classes we add evenly spaced interval bounds in between the extreme bounds. For binary classification ($k$\thinspace=\thinspace2) we consider the hypo/hyper definitions from \cite{kwon2021artificial}\footnote{Calcium: 2.0/2.75; potassium: 3.5/5.5; creatinine: 3.5/5.3; sodium: 130/150. Values in mmol/l. For creatinine we default to $\mu\pm2\sigma$ as \cite{kwon2021artificial} do not consider creatinine.}. For each prediction in the interval, we compute Receiver-Operating-Characteristics (ROC) curves for the classification $p(y \leq i),\,i=1,\dots,k$, leading to $k-1$ individual curves. 
    We compare classification and ordinal regression in the simplified setting with discretized targets $y$. We consider increasingly fine-grained predictions by varying the number of intervals $k$. The class intervals are defined for each electrolyte separately: For $k$\thinspace=\thinspace3 classes, we define the lower and upper interval bounds by $\mu \pm 2 \sigma$. For $k$\thinspace$>$\thinspace3 classes we add evenly spaced interval bounds in between the extreme bounds. For binary classification ($k$\thinspace=\thinspace2) we consider the hypo/hyper definitions from \cite{kwon2021artificial}\footnote{Calcium: 2.0/2.75; potassium: 3.5/5.5; creatinine: 3.5/5.3; sodium: 130/150. Values in mmol/l. For creatinine we default to $\mu\pm2\sigma$ as \cite{kwon2021artificial} do not consider creatinine.}. For evaluation, we compute Receiver-Operating-Characteristic (ROC) curves for the cumulative classification $p(\tilde{y} \leq i),\,i=1,\dots,k$, leading to $k-1$ individual curves.

    % \cref{figure:ROC_Classification} shows the area under the macro averaged ROC (AUmROC) for different number of classes. In general, calcium performs worse than potassium, highlighting further the difficulty of predicting calcium. When increasing the number of classes, we observe a drop in AUmROC which implies that fine-grained predictions are increasingly difficult. \hl{This effect is stronger for calcium, (which strengthens our previous observation from Fig XX %\cref{figure:Regression}
    % that calcium blabla (tie back to former obs)) indicating that there is not sufficient information in the data to predict continuous values and perform regression for calcium}. Comparing classification against ordinal regression, the latter decreases less in AUmROC for more classes, demonstrating that ordinal regression can improve classification performance. \hl{This offers a possible remedy for problems where regression is challenging (be more specific here)}.
    
   \begin{figure}[t]
        \centering
        \includegraphics[width=\columnwidth]{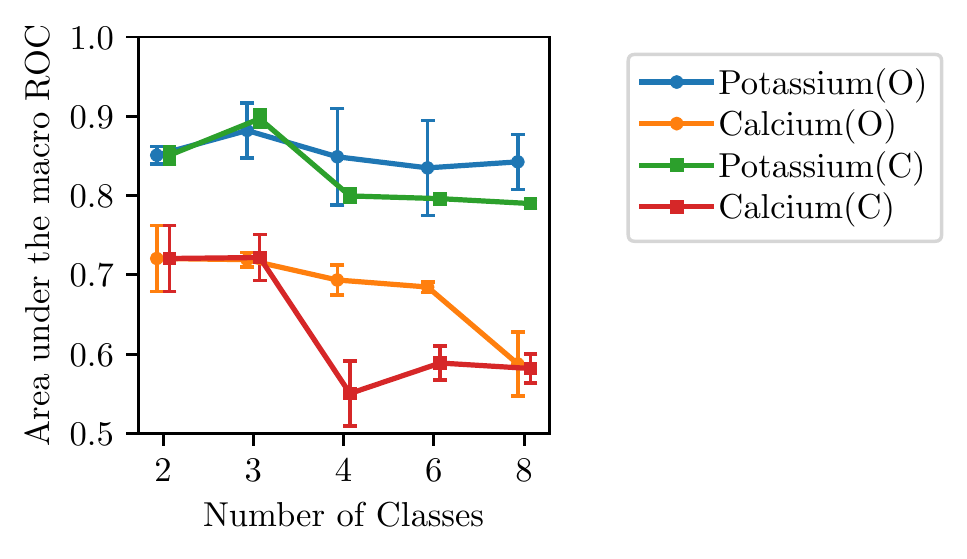}
        \caption{\textbf{Macro ROC for varying number of classes}:
        O: Ordinal regression; C: classification models. For 2 classes we average the hypo and hyper results (see \Cref{table:AUC_2}).}
        \label{figure:ROC_Classification}\vspace{0.0mm}
    \end{figure}

    % \setlength{\tabcolsep}{8pt}
    % \begin{table}[t]
    %     \centering
    %     % \normalsize
    %     \caption{
    %     \textbf{Binary classification AUROC.}}
    %     \begin{tabular}{l|r|ll}
    %         \multicolumn{2}{l}{\textbf{Potassium}} & \multicolumn{2}{l}{(Bounds 3.5/5.5 mmol/l)} \\
    %         \toprule
    %          & $n$ Data & Hypo & Hyper \\
    %         \midrule
    %         ours & 290\thinspace889 & 0.809 (0.003) & 0.892 (0.009)\\
    %         \cite{lin2020deep} & 66\thinspace321 & 0.926 & 0.958 \\
    %         \cite{galloway2019development} & 2\thinspace835\thinspace059 & N/A & 0.865 \\
    %         \cite{kwon2021artificial} & 83\thinspace449 & 0.866 & 0.945 \\
    %         \bottomrule\addlinespace[\belowrulesep]
    %         \multicolumn{2}{l}{\textbf{Calcium}} & \multicolumn{2}{l}{(Bounds 2.0/2.75 mmol/l)}\\
    %         \toprule
    %         ours & 125\thinspace970 & 0.779 (0.012) & 0.660 (0.036) \\
    %         \cite{kwon2021artificial} & 83\thinspace449 & 0.901 & 0.901 \\
    %         \bottomrule
    %     \end{tabular}
    %     \label{table:AUC_2}\vspace{-1.5mm}
    % \end{table}
    % \setlength{\tabcolsep}{6pt}
    \setlength{\tabcolsep}{8pt}
    \begin{table}[t]
        \centering
        % \normalsize
        \caption{
        \textbf{Binary classification AUROC.}}
        \begin{tabular}{l|r|ll}
            \multicolumn{2}{l}{\textbf{Potassium}} & \multicolumn{2}{l}{(Bounds 3.5\thinspace/\thinspace5.5 mmol/l)} \\
            \toprule
             & $n$ Data & Hypo & Hyper \\
            \midrule
            ours & 290\thinspace889 & 0.809 (0.003) & 0.892 (0.009)\\
            \cite{lin2020deep} & 66\thinspace321 & 0.926 & 0.958 \\
            \cite{galloway2019development} & 2\thinspace835\thinspace059 & N/A & 0.865 \\
            \cite{kwon2021artificial} & 83\thinspace449 & 0.866 & 0.945 \\
            \bottomrule\addlinespace[\belowrulesep]
            \multicolumn{2}{l}{\textbf{Calcium}} & \multicolumn{2}{l}{(Bounds 2.0\thinspace/\thinspace2.75 mmol/l)}\\
            \toprule
            ours & 125\thinspace970 & 0.779 (0.012) & 0.660 (0.036) \\
            \cite{kwon2021artificial} & 83\thinspace449 & 0.901 & 0.905 \\
            \bottomrule
        \end{tabular}
        \label{table:AUC_2}\vspace{-1.5mm}
    \end{table}
    \setlength{\tabcolsep}{6pt}
    
    \Cref{figure:ROC_Classification} shows the area under the macro averaged ROC (AUmROC) for different number of classes $k$. The AUmROC simply averages the obtained $k-1$ AUROC values. For all $k$, the prediction performance on calcium is worse than on potassium. When increasing the number of classes, we observe a drop in AUmROC which implies that fine-grained predictions are increasingly difficult. This effect is also stronger for calcium. Together with \cref{figure:Regression}, these results clearly suggest that accurate prediction of concentration levels is inherently more difficult for calcium than for potassium. Comparing classification against ordinal regression in \cref{figure:ROC_Classification}, the latter decreases less in AUmROC for more classes. In this discretized regression setting, ordinal regression can thus improve performance compared to standard classification models.

    We now convert the class predictions into electrolyte concentration levels by mapping to the mean of the predicted class interval, and compute the error to the continuous targets. The results in \cref{figure:AppendixMaePotassiumCalcium} show that the MAE decreases with more classes for potassium but stays mostly constant for calcium. However, the MAE is never lower than the corresponding direct regression MAE from \Cref{table:Regression}. While discretization thus can lead to good \emph{classification} performance, it does not help solve the original problem of predicting continuous concentration levels.
    
    For binary classification, we compare with results from literature in \Cref{table:AUC_2}. The comparisons are not entirely fair since data collection and dataset size is different between all works. For potassium, our model reaches a slightly lower AUROC for both imbalances than 2 out of 3 works from literature. For calcium, Kwon et al.~\cite{kwon2021artificial} reach a significantly higher AUROC.% than our method.  %Possible reasons are the usage of a different dataset, especially in the case of calcium where there is only one other work to compare to.

\subsection{Probabilistic Regression}
   \begin{figure}[t]
        \centering
        \includegraphics[width=0.9\columnwidth]{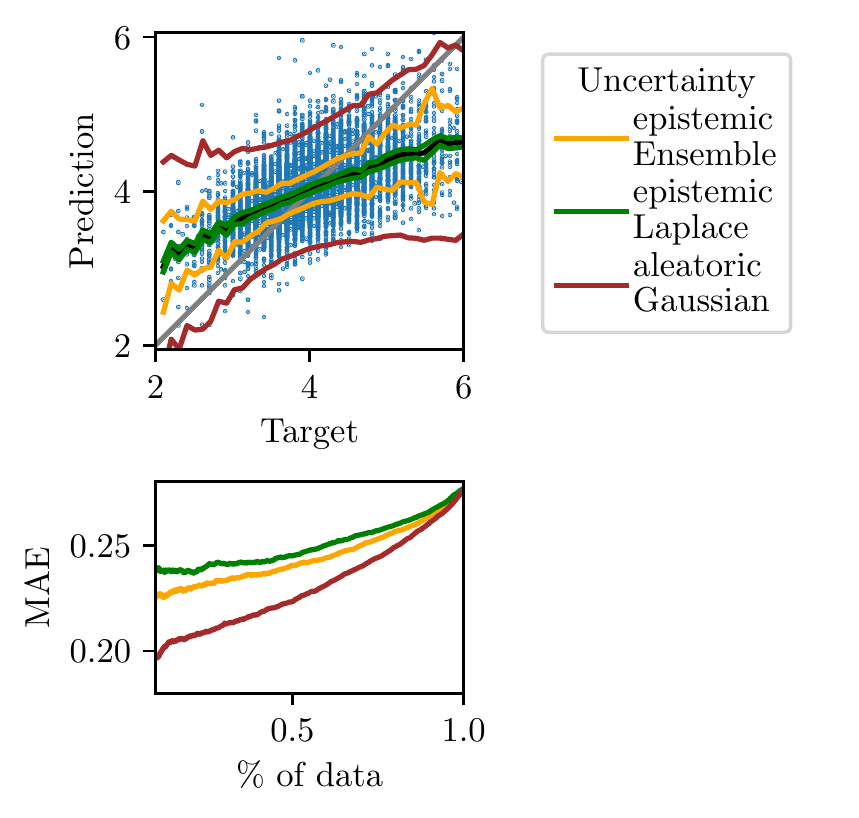}\vspace{-2.0mm} %0.8 \textwidth wide
        \caption{\textbf{Regression uncertainty}: \textit{Top}: Prediction vs target plot as in \cref{figure:Regression}. Black line indicates mean prediction $\mu$. The coloured lines show $\mu \pm 2 \sigma$ for different types of uncertainties.\! \textit{Bottom}: Sparsification plot.}
        \label{figure:RegressionUncertainty}\vspace{-1.5mm}
    \end{figure}

    Here, we focus on potassium as the only electrolyte for which direct regression learns a clear relationship between inputs and targets. \Cref{figure:RegressionUncertainty} (top) shows the three uncertainty estimates defined in \Cref{section:methods:model_training}, for different target levels. Aleatoric Gaussian uncertainty fits the noise in the predictions quite well, since it increases towards the extremes, where predictions become noisy and the error increases. In comparison, both epistemic variances are smaller, which is expected due to the large size of the underlying training dataset. Epistemic Laplace is the smallest and almost constant. Epistemic from the ensemble increases towards the extreme values, similar to the aleatoric uncertainty.
    
    Meaningful uncertainty estimates should correlate with the error -- predictions with high error should also have high uncertainty. The sparsification plot in \cref{figure:RegressionUncertainty} (bottom) shows that removing the most uncertain points lowers the MAE monotonically, as expected. This effect is strongest for aleatoric Gaussian uncertainty. However, the calibration plot in \cref{figure:AppendixCalibration} shows that the uncertainties are not particularly well-calibrated. Aleatoric Gaussian uncertainty has the highest correlation with the MSE, see \Cref{table:AppendixCorrelation}. The correlation can be increased by adding epistemic ensemble uncertainty. 
    
    To further measure uncertainty, we perform OOD experiments similar to  \cite{xia2021benchmarking}. We first add Gaussian noise to the ECG traces, controlling the strength with the signal-to-noise ratio (SNR). In \Cref{table:OOD} we observe that with increasing noise both the MAE and all uncertainty measures rise. This indicates that each uncertainty by itself is a useful indicator for this kind of OOD data. As a second experiment, we randomly mask a proportion of each ECG trace. The results for this experiment are provided in \Cref{table:appendixOOD} and indicate that this type of OOD data is not detected by our uncertainty quantification.    
    % \hl{While the MAE rises slightly with increased masking, the uncertainty measures stay mostly constant or even decrease. They are thus not very useful indicators for this specific type of OOD data}. \hl{(Too long? Too negative?)}

% \Cref{table:appendixOOD} lists the results of the OOD experiments. While the experiments for the SNR are expected (larger MAE and uncertainties for lower SNR), the results for masking are not as clear. While the MAE still increases, notably especially the epistemic ensemble uncertainty decreases. This means that there is less variance in the mean predictions between the different ensemble members.
    
    \setlength{\tabcolsep}{8pt}  
    \begin{table}[t]
        \centering
        % \normalsize
        \caption{\textbf{OOD experiment}: AG: Aleatoric Gaussian; EE: Epistemic ensemble; EL: Epistemic Laplace. }
        % \resizebox{0.95\columnwidth}{!}{
        \begin{tabular}{l|lll}%
            \toprule
             & Baseline & SNR 10 & SNR 1 \\
            \midrule
            MAE & 0.304 (0.021) & 0.330 (0.016) &  0.368 (0.026) \\
            \midrule
            AG & 0.389 (0.012) & 0.399 (0.012) & 0.480 (0.078) \\
            EE & 0.121 (0.048) & 0.149 (0.041) & 0.184 (0.075) \\
            EL & 0.022 (0.003) & 0.028 (0.009) & 0.049 (0.031) \\
            \bottomrule
        \end{tabular}
        % }
        \label{table:OOD}\vspace{-1.5mm}
    \end{table}
    \setlength{\tabcolsep}{6pt}  
    
    %shows that for Gaussian noise with decreasing SNR, both Laplace and Ensemble uncertainty increases compared to the baseline. For masking, the Ensemble method however decreases in uncertainty the larger the mask, which in turn means that the prediction between the different ensemble models have less variance. For Laplace on the other hand the variance increases initially for small masks, but also decreases for larger masks. 
    %As a second experiment, we mask a variable sized block in time for all leads. \Cref{table:OOD} shows that with increasing shift, the model performance measured by MSE drops. 

    % \addtolength{\tabcolsep}{-2pt} 
    % \begin{table*}[]
    % \centering
    % \caption{\textbf{OOD experiments}: SNR refers to the Signal-to-Noise ratio for Gaussian noise, Mask X refers to the percent of the data masked. \hl{Make this table look nicer---have to write manually! TODO!} \hl{Reduce the number of significant digits, TODO!} \hl{decide on what rows to keep}}
    % \resizebox{\textwidth}{!}{
    % {\begin{tabular}{p{2cm}|p{2.5cm}|p{2.5cm}p{2.5cm}p{2.5cm}p{2.5cm}}%
    %     \toprule
    %     Experiment & Ensemble MAE & Aleatoric Gaussian & Epistemic ensemble & Epistemic Laplace & Epistemic direct Regression\\
    %     \midrule
    %     \csvreader[head to column names]{tables/ood_results.csv}{}% use head of csv as column names
    %     {\\ \experiment & \MAEMean (\MAEStd) & \GaussianMean (\GaussianStd) &  \EnsembleMean (\EnsembleStd) & \LaplaceMean(\LaplaceStd) & \RegressionEnsemble}\\% specify your columns here
    %     \bottomrule
    % \end{tabular}}
    % }
    % \label{table:OOD}
    % \end{table*}
    % \addtolength{\tabcolsep}{-1pt} 

    \subsection{Discussion}
    
    The varying performance of our models across electrolytes, see for instance \cref{figure:Regression}, requires discussion. 
    % explanation why calcium does not work well
    The difficulty in predicting calcium levels in contrast to potassium levels can be justified. There is a known relationship between both electrolyte levels and a change of the ECG, which for calcium is manifested mainly in a change of the QT interval \cite{Gardner2014,chorin2016electrocardiographic}. However, the range of values for calcium is very narrow since it is tightly regulated by many mechanisms in the human body (extreme values can lead to death). In our dataset, about 95\% of the values are in the range $2.29\pm0.26$, see \Cref{table:dataset}. The electropysiological manifestation of this change in concentration level could be negligible as Fig.~2 in \cite{pilia2019ecg} indicates. Further, the calcium dataset size is less than half that of potassium, thus the number of patients with extreme calcium values could be insufficient to predict those reliably.
    
    % most effects are manifested in potassium, hence potassium prediction works well
    The electrophysiological effects of potassium, calcium, sodium and creatinine are closely interlinked. For example, extracellular hypo- and hyperkalemia (potassium) levels promote cardiac arrhythmias, partly because of direct potassium effects, and partly because the intracellular balances of potassium, sodium and calcium are linked. Thus, hypo- and hyperkalemia directly impact sodium and calcium balances. %Hypokalemia also causes intracellular sodium and calcium accumulation.% , which promotes arrhythmias. Hyperkalemia on the other hand promotes conduction block and reentry and may also promote reentry and arrhythmias. 
    Finally, creatinine levels can signify renal disease that can lead to hyperkalemia. This complex relationship between the studied electrolytes in addition with the significant better regression results for potassium could indicate that the summed electrophysiological effects are most tightly linked to potassium concentrations. However, due to the known connection of calcium to the ECG, it is unclear if the combined electrophysiological effects could also be linked to calcium if we had a similar dataset size as for potassium.
    
    % We observe that especially the potassium levels can be predicted quite accurately; this is clinically justified. The electrophysiological effects of potassium, calcium, sodium and creatinine are closely interlinked. For example, extracellular hypo- and hyperkalemia (potassium) levels promote cardiac arrhythmias, partly because of direct potassium effects, and partly because the intracellular balances of potassium, sodium and calcium are linked. Thus, hypo- and hyperkalemia directly impact sodium and calcium balances. Hypokalemia also causes intracellular sodium and calcium accumulation, which promotes arrhythmias. Hyperkalemia on the other hand promotes conduction block and reentry and may also promote reentry and arrhythmias. Finally, creatinine levels can signify renal disease that can lead to hyperkalemia. Given these complex relationships, this study could indicate that the summed cardiac electrophysiological effects are most tightly linked to potassium concentrations. 

    % A possible connection to calcium could be drawn if the model accuracy increases given a similarly large dataset size as for potassium.

\section{Conclusion}
\label{section:Conclusion}

We trained deep models for direct regression of continuous electrolyte concentration levels from ECGs. While the model for potassium performed quite well, it struggled for the three other electrolytes. Simplifying the problem to binary classification, of clinically critical low or high levels, indicated that also those electrolytes for which the direct regression model struggled can achieve good classification performance. Defining more classes, for increasingly fine-grained predictions, we observed a sharp performance drop for electrolytes other than potassium. Our results thus strongly suggest that accurate ECG-based prediction of concentration levels is inherently more difficult for some electrolytes than for others. Future work should study this problem for an even larger set of electrolytes, and explore the possibility of combined models for all electrolytes due to their medial interconnection.

We also extended our deep direct regression model to the probabilistic regression approach and carefully analyzed the resulting uncertainty estimates. While especially the aleatoric uncertainty demonstrated potential practical usefulness e.g. via sparsification, the uncertainty estimates are not particularly well-calibrated. To achieve the ultimate goal of accurate, reliable and clinically useful prediction of electrolyte concentration levels, future work investigating possible approaches for improved uncertainty calibration is thus required.

\clearpage
\bibliographystyle{IEEEtran}
\bibliography{references.bib} % IEEEabrv

\begin{thebibliography}{10}
\providecommand{\url}[1]{#1}
\csname url@samestyle\endcsname
\providecommand{\newblock}{\relax}
\providecommand{\bibinfo}[2]{#2}
\providecommand{\BIBentrySTDinterwordspacing}{\spaceskip=0pt\relax}
\providecommand{\BIBentryALTinterwordstretchfactor}{4}
\providecommand{\BIBentryALTinterwordspacing}{\spaceskip=\fontdimen2\font plus
\BIBentryALTinterwordstretchfactor\fontdimen3\font minus
  \fontdimen4\font\relax}
\providecommand{\BIBforeignlanguage}[2]{{%
\expandafter\ifx\csname l@#1\endcsname\relax
\typeout{** WARNING: IEEEtran.bst: No hyphenation pattern has been}%
\typeout{** loaded for the language `#1'. Using the pattern for}%
\typeout{** the default language instead.}%
\else
\language=\csname l@#1\endcsname
\fi
#2}}
\providecommand{\BIBdecl}{\relax}
\BIBdecl

\bibitem{EDELMAN1959256}
I.~Edelman and J.~Leibman, ``Anatomy of body water and electrolytes,''
  \emph{The American Journal of Medicine}, vol.~27, no.~2, pp. 256--277, 1959.

\bibitem{CARLSON2008529}
G.~P. Carlson and M.~Bruss, ``Chapter 17 - fluid, electrolyte, and acid-base
  balance,'' in \emph{Clinical Biochemistry of Domestic Animals (Sixth
  Edition)}, sixth edition~ed., J.~J. Kaneko, J.~W. Harvey, and M.~L. Bruss,
  Eds.\hskip 1em plus 0.5em minus 0.4em\relax Academic Press, 2008, pp.
  529--559.

\bibitem{paice1986record}
B.~Paice, K.~Paterson, F.~Onyanga-Omara, T.~Donnelly, J.~Gray, and D.~Lawson,
  ``Record linkage study of hypokalaemia in hospitalized patients.''
  \emph{Postgraduate medical journal}, vol.~62, no. 725, pp. 187--191, 1986.

\bibitem{el2011electrolyte}
N.~El-Sherif and G.~Turitto, ``Electrolyte disorders and arrhythmogenesis,''
  \emph{Cardiology journal}, vol.~18, no.~3, pp. 233--245, 2011.

\bibitem{fisch1973relation}
C.~Fisch, ``Relation of electrolyte disturbances to cardiac arrhythmias,''
  \emph{Circulation}, vol.~47, no.~2, pp. 408--419, 1973.

\bibitem{SURAWICZ1967814}
B.~Surawicz, ``Relationship between electrocardiogram and electrolytes,''
  \emph{American Heart Journal}, vol.~73, no.~6, pp. 814--834, 1967.

\bibitem{macfarlane2005university}
P.~Macfarlane, B.~Devine, and E.~Clark, ``The university of {G}lasgow
  ({U}ni-{G}) {ECG} analysis program,'' in \emph{Computers in Cardiology,
  2005}, 2005, pp. 451--454.

\bibitem{rajpurkar2017cardiologist}
P.~Rajpurkar, A.~Y. Hannun, M.~Haghpanahi, C.~Bourn, and A.~Y. Ng,
  ``Cardiologist-level arrhythmia detection with convolutional neural
  networks,'' \emph{arXiv preprint arXiv:1707.01836}, 2017.

\bibitem{hannun2019cardiologist}
A.~Y. Hannun, P.~Rajpurkar, M.~Haghpanahi, G.~H. Tison, C.~Bourn, M.~P.
  Turakhia, and A.~Y. Ng, ``Cardiologist-level arrhythmia detection and
  classification in ambulatory electrocardiograms using a deep neural
  network,'' \emph{Nature medicine}, vol.~25, no.~1, pp. 65--69, 2019.

\bibitem{ribeiro2020automatic}
A.~H. Ribeiro, M.~H. Ribeiro, G.~M. Paix{\~a}o, D.~M. Oliveira, P.~R. Gomes,
  J.~A. Canazart, M.~P. Ferreira, C.~R. Andersson, P.~W. Macfarlane,
  W.~Meira~Jr \emph{et~al.}, ``Automatic diagnosis of the 12-lead {ECG} using a
  deep neural network,'' \emph{Nature communications}, vol.~11, no.~1, pp.
  1--9, 2020.

\bibitem{gustafsson_nstemi}
S.~Gustafsson, D.~Gedon, E.~Lampa, A.~H. Ribeiro, M.~J. Holzmann, T.~B.
  Sch\"{o}n, and J.~Sundstr\"{o}m, ``Development and validation of deep
  learning {ECG}-based prediction of myocardial infarction in emergency
  department patients,'' \emph{Scientific Reports}, vol.~12, no.~1, 2022.

\bibitem{raghunath2020prediction}
S.~Raghunath, A.~E. Ulloa~Cerna, L.~Jing, D.~P. VanMaanen, J.~Stough, D.~N.
  Hartzel, J.~B. Leader, H.~L. Kirchner, M.~C. Stumpe, A.~Hafez \emph{et~al.},
  ``Prediction of mortality from 12-lead electrocardiogram voltage data using a
  deep neural network,'' \emph{Nature medicine}, vol.~26, no.~6, pp. 886--891,
  2020.

\bibitem{lima2021deep}
E.~M. Lima, A.~H. Ribeiro, G.~M. Paix{\~a}o, M.~H. Ribeiro, M.~M. Pinto-Filho,
  P.~R. Gomes, D.~M. Oliveira, E.~C. Sabino, B.~B. Duncan, L.~Giatti
  \emph{et~al.}, ``Deep neural network-estimated electrocardiographic age as a
  mortality predictor,'' \emph{Nature communications}, vol.~12, no.~1, pp.
  1--10, 2021.

\bibitem{attia_artificial_2019}
Z.~I. Attia, P.~A. Noseworthy, F.~Lopez-Jimenez, S.~J. Asirvatham, A.~J.
  Deshmukh, B.~J. Gersh, R.~E. Carter, X.~Yao, A.~A. Rabinstein, B.~J.
  Erickson, S.~Kapa, and P.~A. Friedman, ``An artificial intelligence-enabled
  {ECG} algorithm for the identification of patients with atrial fibrillation
  during sinus rhythm: a retrospective analysis of outcome prediction,''
  \emph{The Lancet}, 2019.

\bibitem{biton_atrial_2021}
S.~Biton, S.~Gendelman, A.~H. Ribeiro, G.~Miana, C.~Moreira, A.~L.~P. Ribeiro,
  and J.~A. Behar, ``Atrial fibrillation risk prediction from the 12-lead {ECG}
  using digital biomarkers and deep representation learning,'' \emph{European
  Heart Journal - Digital Health}, 2021.

\bibitem{attia_screening_2019}
Z.~I. Attia, S.~Kapa, F.~Lopez-Jimenez, P.~M. McKie, D.~J. Ladewig, G.~Satam,
  P.~A. Pellikka, M.~Enriquez-Sarano, P.~A. Noseworthy, T.~M. Munger, S.~J.
  Asirvatham, C.~G. Scott, R.~E. Carter, and P.~A. Friedman, ``Screening for
  cardiac contractile dysfunction using an artificial intelligence–enabled
  electrocardiogram,'' \emph{Nature Medicine}, vol.~25, no.~1, pp. 70--74,
  2019.

\bibitem{kwon2021artificial}
J.-m. Kwon, M.-S. Jung, K.-H. Kim, Y.-Y. Jo, J.-H. Shin, Y.-H. Cho, Y.-J. Lee,
  J.-H. Ban, K.-H. Jeon, S.~Y. Lee \emph{et~al.}, ``Artificial intelligence for
  detecting electrolyte imbalance using electrocardiography,'' \emph{Annals of
  Noninvasive Electrocardiology}, vol.~26, no.~3, p. e12839, 2021.

\bibitem{gast2018lightweight}
J.~Gast and S.~Roth, ``Lightweight probabilistic deep networks,'' in
  \emph{Proceedings of the IEEE Conference on Computer Vision and Pattern
  Recognition (CVPR)}, 2018, pp. 3369--3378.

\bibitem{varamesh2020mixture}
A.~Varamesh and T.~Tuytelaars, ``Mixture dense regression for object detection
  and human pose estimation,'' in \emph{Proceedings of the IEEE/CVF Conference
  on Computer Vision and Pattern Recognition (CVPR)}, 2020, pp.
  13\,086--13\,095.

\bibitem{xiao2018simple}
B.~Xiao, H.~Wu, and Y.~Wei, ``Simple baselines for human pose estimation and
  tracking,'' in \emph{Proceedings of the European Conference on Computer
  Vision (ECCV)}, 2018, pp. 466--481.

\bibitem{ruiz2018fine}
N.~Ruiz, E.~Chong, and J.~M. Rehg, ``Fine-grained head pose estimation without
  keypoints,'' in \emph{Proceedings of the IEEE Conference on Computer Vision
  and Pattern Recognition (CVPR) Workshops}, 2018, pp. 2074--2083.

\bibitem{gustafsson2020energy}
F.~K. Gustafsson, M.~Danelljan, G.~Bhat, and T.~B. Sch{\"o}n, ``Energy-based
  models for deep probabilistic regression,'' in \emph{Proceedings of the
  European Conference on Computer Vision (ECCV)}, 2020.

\bibitem{lathuiliere2019comprehensive}
S.~Lathuili{\`e}re, P.~Mesejo, X.~Alameda-Pineda, and R.~Horaud, ``A
  comprehensive analysis of deep regression,'' \emph{IEEE Transactions on
  Pattern Analysis and Machine Intelligence (TPAMI)}, 2019.

\bibitem{frohnert1970statistical}
P.~P. Frohnert, E.~R. Gluliani, M.~Friedberg, W.~J. Johnson, and W.~N. Tauxe,
  ``Statistical investigation of correlations between serum potassium levels
  and electrocardiographic findings in patients on intermittent hemodialysis
  therapy,'' \emph{Circulation}, vol.~41, no.~4, pp. 667--676, 1970.

\bibitem{velagapudi2017computer}
V.~Velagapudi, J.~C. O'Horo, A.~Vellanki, S.~P. Baker, R.~Pidikiti, J.~S.
  Stoff, and D.~A. Tighe, ``Computer-assisted image processing 12 lead {ECG}
  model to diagnose hyperkalemia,'' \emph{Journal of electrocardiology},
  vol.~50, no.~1, pp. 131--138, 2017.

\bibitem{galloway2019development}
C.~D. Galloway, A.~V. Valys, J.~B. Shreibati, D.~L. Treiman, F.~L. Petterson,
  V.~P. Gundotra, D.~E. Albert, Z.~I. Attia, R.~E. Carter, S.~J. Asirvatham
  \emph{et~al.}, ``Development and validation of a deep-learning model to
  screen for hyperkalemia from the electrocardiogram,'' \emph{JAMA cardiology},
  vol.~4, no.~5, pp. 428--436, 2019.

\bibitem{lin2020deep}
C.-S. Lin, C.~Lin, W.-H. Fang, C.-J. Hsu, S.-J. Chen, K.-H. Huang, W.-S. Lin,
  C.-S. Tsai, C.-C. Kuo, T.~Chau \emph{et~al.}, ``A deep-learning algorithm
  ({ECG}12{N}et) for detecting hypokalemia and hyperkalemia by
  electrocardiography: algorithm development,'' \emph{JMIR medical
  informatics}, vol.~8, no.~3, p. e15931, 2020.

\bibitem{kendall2017uncertainties}
A.~Kendall and Y.~Gal, ``What uncertainties do we need in {B}ayesian deep
  learning for computer vision?'' in \emph{Advances in Neural Information
  Processing Systems (NeurIPS)}, 2017, pp. 5574--5584.

\bibitem{gal2016thesis}
Y.~Gal, ``Uncertainty in deep learning,'' Ph.D. dissertation, University of
  Cambridge, 2016.

\bibitem{lakshminarayanan2017simple}
B.~Lakshminarayanan, A.~Pritzel, and C.~Blundell, ``Simple and scalable
  predictive uncertainty estimation using deep ensembles,'' in \emph{Advances
  in Neural Information Processing Systems (NeurIPS)}, 2017, pp. 6402--6413.

\bibitem{neal1995bayesian}
R.~M. Neal, ``{B}ayesian learning for neural networks,'' Ph.D. dissertation,
  University of Toronto, 1995.

\bibitem{dietterich2000ensemble}
T.~G. Dietterich, ``Ensemble methods in machine learning,'' in
  \emph{International workshop on multiple classifier systems}.\hskip 1em plus
  0.5em minus 0.4em\relax Springer, 2000, pp. 1--15.

\bibitem{ovadia2019can}
Y.~Ovadia, E.~Fertig, J.~Ren, Z.~Nado, D.~Sculley, S.~Nowozin, J.~Dillon,
  B.~Lakshminarayanan, and J.~Snoek, ``Can you trust your model\textquotesingle
  s uncertainty? {E}valuating predictive uncertainty under dataset shift,'' in
  \emph{Advances in Neural Information Processing Systems (NeurIPS)}, vol.~32,
  2019.

\bibitem{gustafsson2020evaluating}
F.~K. Gustafsson, M.~Danelljan, and T.~B. Schön, ``Evaluating scalable
  bayesian deep learning methods for robust computer vision,'' in
  \emph{Proceedings of the IEEE/CVF conference on computer vision and pattern
  recognition workshops}, 2020, pp. 318--319.

\bibitem{mackay1992bayesian}
D.~J. MacKay, ``Bayesian interpolation,'' \emph{Neural computation}, vol.~4,
  no.~3, pp. 415--447, 1992.

\bibitem{kristiadi2020being}
A.~Kristiadi, M.~Hein, and P.~Hennig, ``Being {B}ayesian, even just a bit,
  fixes overconfidence in {R}e{LU} networks,'' in \emph{International
  conference on machine learning}.\hskip 1em plus 0.5em minus 0.4em\relax PMLR,
  2020, pp. 5436--5446.

\bibitem{daxberger2021laplace}
E.~Daxberger, A.~Kristiadi, A.~Immer, R.~Eschenhagen, M.~Bauer, and P.~Hennig,
  ``Laplace redux--effortless {B}ayesian deep learning,'' \emph{arXiv preprint
  arXiv:2106.14806}, 2021.

\bibitem{torgo1996regression}
L.~Torgo and J.~Gama, ``Regression by classification,'' in \emph{Brazilian
  symposium on artificial intelligence}.\hskip 1em plus 0.5em minus 0.4em\relax
  Springer, 1996, pp. 51--60.

\bibitem{cao2020rank}
W.~Cao, V.~Mirjalili, and S.~Raschka, ``Rank consistent ordinal regression for
  neural networks with application to age estimation,'' \emph{Pattern
  Recognition Letters}, vol. 140, pp. 325--331, 2020.

\bibitem{Alkmim2012}
M.~B. Alkmim, R.~M. Figueira, M.~S. Marcolino, C.~S. Cardoso, M.~P. de~Abreu,
  L.~R. Cunha, D.~F. da~Cunha, A.~P. Antunes, A.~G. de~A~Resende, E.~S.
  Resende, and A.~L.~P. Ribeiro, ``Improving patient access to specialized
  health care: the telehealth network of minas gerais, brazil,'' \emph{Bulletin
  of the World Health Organization}, vol.~90, no.~5, pp. 373--378, 2012.

\bibitem{hendrycks2018benchmarking}
D.~Hendrycks and T.~Dietterich, ``Benchmarking neural network robustness to
  common corruptions and perturbations,'' in \emph{International Conference on
  Learning Representations (ICLR)}, 2019.

\bibitem{koh2021wilds}
P.~W. Koh, S.~Sagawa, H.~Marklund, S.~M. Xie, M.~Zhang, A.~Balsubramani, W.~Hu,
  M.~Yasunaga, R.~L. Phillips, I.~Gao \emph{et~al.}, ``Wilds: A benchmark of
  in-the-wild distribution shifts,'' in \emph{International Conference on
  Machine Learning (ICML)}.\hskip 1em plus 0.5em minus 0.4em\relax PMLR, 2021,
  pp. 5637--5664.

\bibitem{attia2016novel}
Z.~I. Attia, C.~V. DeSimone, J.~J. Dillon, Y.~Sapir, V.~K. Somers, J.~L. Dugan,
  C.~J. Bruce, M.~J. Ackerman, S.~J. Asirvatham, B.~L. Striemer \emph{et~al.},
  ``Novel bloodless potassium determination using a signal-processed
  single-lead {ECG},'' \emph{Journal of the American heart Association},
  vol.~5, no.~1, p. e002746, 2016.

\bibitem{corsi2017noninvasive}
C.~Corsi, M.~Cortesi, G.~Callisesi, J.~De~Bie, C.~Napolitano, A.~Santoro,
  D.~Mortara, and S.~Severi, ``Noninvasive quantification of blood potassium
  concentration from {ECG} in hemodialysis patients,'' \emph{Scientific
  Reports}, vol.~7, no.~1, pp. 1--10, 2017.

\bibitem{xia2021benchmarking}
T.~Xia, J.~Han, and C.~Mascolo, ``Benchmarking uncertainty qualification on
  biosignal classification tasks under dataset shift,'' \emph{arXiv preprint
  arXiv:2112.09196}, 2021.

\bibitem{friedman2001greedy}
J.~H. Friedman, ``Greedy function approximation: a gradient boosting machine,''
  \emph{Annals of statistics}, pp. 1189--1232, 2001.

\bibitem{breiman2001random}
L.~Breiman, ``Random forests,'' \emph{Machine learning}, vol.~45, no.~1, pp.
  5--32, 2001.

\bibitem{Ebrahimi_2020}
Z.~Ebrahimi, M.~Loni, M.~Daneshtalab, and A.~Gharehbaghi, ``A review on deep
  learning methods for {ECG} arrhythmia classification,'' \emph{Expert Systems
  with Applications: X}, vol.~7, p. 100033, 2020.

\bibitem{Xiong_2022}
P.~Xiong, S.~M.-Y. Lee, and G.~Chan, ``Deep learning for detecting and locating
  myocardial infarction by electrocardiogram: A literature review,''
  \emph{Frontiers in Cardiovascular Medicine}, vol.~9, 2022.

\bibitem{paszke2019pytorch}
A.~Paszke, S.~Gross, F.~Massa, A.~Lerer, J.~Bradbury, G.~Chanan, T.~Killeen,
  Z.~Lin, N.~Gimelshein, L.~Antiga \emph{et~al.}, ``Pytorch: An imperative
  style, high-performance deep learning library,'' \emph{Advances in neural
  information processing systems}, vol.~32, 2019.

\bibitem{Gardner2014}
J.~D. Gardner, J.~B. Calkins, and G.~E. Garrison, ``{ECG} diagnosis: The effect
  of ionized serum calcium levels on electrocardiogram,'' \emph{The Permanente
  Journal}, vol.~18, no.~1, 2014.

\bibitem{chorin2016electrocardiographic}
E.~Chorin, R.~Rosso, and S.~Viskin, ``Electrocardiographic manifestations of
  calcium abnormalities,'' \emph{Annals of Noninvasive Electrocardiology: The
  Official Journal of the International Society for Holter and Noninvasive
  Electrocardiology, Inc}, vol.~21, no.~1, p.~7, 2016.

\bibitem{pilia2019ecg}
N.~Pilia, M.~H. Mesa, O.~D{\"o}ssel, and A.~Loewe, ``{ECG}-based estimation of
  potassium and calcium concentrations: Proof of concept with simulated data,''
  in \emph{2019 41st Annual International Conference of the IEEE Engineering in
  Medicine and Biology Society (EMBC)}.\hskip 1em plus 0.5em minus 0.4em\relax
  IEEE, 2019, pp. 2610--2613.

\bibitem{balci2013general}
A.~K. Balc{\i}, O.~Koksal, A.~Kose, E.~Armagan, F.~Ozdemir, T.~Inal, and
  N.~Oner, ``General characteristics of patients with electrolyte imbalance
  admitted to emergency department,'' \emph{World journal of emergency
  medicine}, vol.~4, no.~2, p. 113, 2013.

\end{thebibliography}

\clearpage
\beginsupplement
\onecolumn
\appendix
\section{Ethical approval}
\label{appendix:Ethics}

% Our study is approved by the relevant ethical review authority. %The relevant information will be provided if the paper is accepted. 
% for camera ready version include below:
\noindent The Ethics Review Board in Stockholm and the Swedish Ethical Review Authority have approved the study (reference numbers 2018/1089-31, 2018/2328-32, 2019-02329, 2019-02339, 2020-01654, 2020-05925, 2021-01668, and 2021-06462).
\subsection{Dataset}
\label{appendix:dataset}

\subsubsection{Clarification of Electrolyte Definitions}

We note that potassium, calcium and sodium are by definition electrolytes but creatinine is an abundant blood biomarker. For the ease of reading on descriptions, we denote creatinine as an electrolyte in this study. The reason to include creatinine is the availability of large amounts of data and the general medical interest in its predictions. In some figures and tables in this appendix, creatinine is denoted ''pcreatinine``.

\subsubsection{Dataset Characteristics}
\label{appendix:dataset char}

%\hl{include "all-comer emergency department patients >=18 years old where the only restriction is that they should have a the blood biomarker and ECG collected within 60m."}

The characteristics of our four datasets are given in \Cref{table:dataset}. More generally a population of emergency room patients with electrolyte imbalances has characteristics as described in \cite{balci2013general}. We include data from all-comer patients to the emergency room with $\geq18$ years old with the only restriction that there is a blood biomarker test and ECG collected within 60 minutes. We have varying number of patients in each dataset because not all electrolyte concentration values are available for all patients. Note that we use an inclusion filter of $\pm60~$minutes between ECG and blood measurement. We can compare our datasets with related work from literature:
\begin{itemize}
    \item \cite{lin2020deep} uses 66\thinspace321 ECG recordings from 40\thinspace180 patients and related potassium concentration in a time frame of $\pm60~$minutes.
    \item \cite{galloway2019development} uses 2\thinspace835\thinspace059 ECG recordings from 787\thinspace661 patients and related potassium concentrations. The authors develop their model on $60\,\%(=449\thinspace380)$ of the patients. All ECGs were recorded within 4~hours before a potassium measurements.
    \item \cite{kwon2021artificial} has 92\thinspace140 patients, whereof 48\thinspace356 patients were used for model development with 83\thinspace449 ECGs. The study considered potassium, sodium and calcium within $\pm30~$minutes of ECG recordings.
\end{itemize}

\begin{figure}[H]
    \centering
    \includegraphics[width=\textwidth]{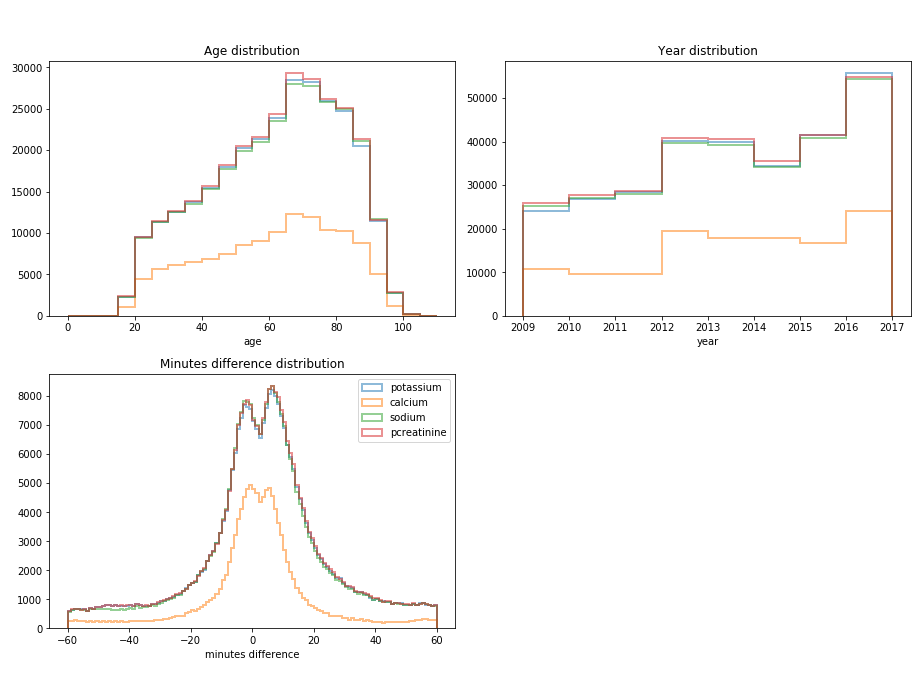}
    \caption{Histogram of meta data age (top left), recording year (top right) and minutes difference between ECG recording and blood measurement (bottom) for our four datasets.}
    \label{fig:dataset hist meta}
\end{figure}

\begin{figure}[H]
    \centering
    \includegraphics[width=0.9\textwidth]{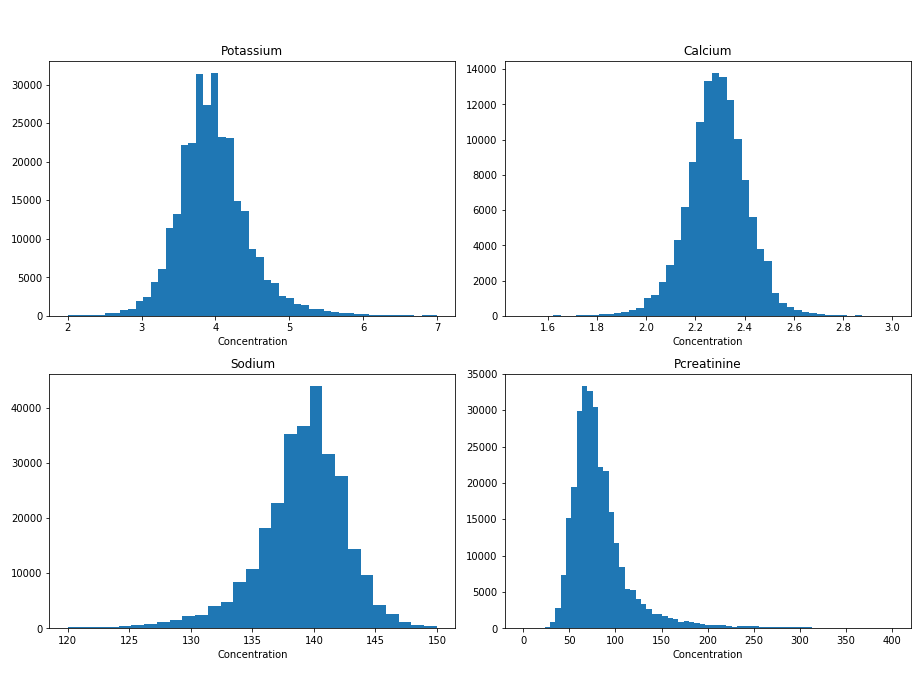}
    \caption{Histogram of the concentration of electrolyte values for our four datasets.}
    \label{fig:dataset hist conc}
\end{figure}

We analysed our datasets in more detail to observe possible causes of errors or shortcuts for our model. In \cref{fig:dataset hist meta} we show histograms of age, recording year and time difference between ECG recording and blood measurement. In \cref{fig:dataset hist conc} we show the distribution of electrolyte concentrations for all four electrolytes, which shows a Normal distribution for all electrolytes except for creatinine which is skewed towards large values. In order to validate our inclusion filter of $\pm60~$minutes, we analyse the concentration of electrolytes vs the time difference and observe no change of concentration value over time. A similar analysis is done for age and sex. Here, we observe that older patients tend to have more extreme electrolyte concentration values for all four electrolytes.
%\Daniel{I did not include those plots for now because it seems redundant, but we can include them of course if we like.}

%The blood measurements were analysed in five laboratories in the Stockholm area. We observed that from the different laboratories there is a difference in time recordings. The origin of this is a difference in summer/winter time from the recording machines. After unsuccessful consultation with the laboratories, we corrected the summer/winter time issue with adding/subtracting one hour to the blood measurement recording of some laboratories such that for all our data points, the median difference between ECG recording and laboratory measurement is zero over the complete year. \hl{stefans comment that the correction is too vague without a plot.}

\subsubsection{Pre-processing}
\label{appendix:dataset prepro}

For the high pass filter to remove the baseline (trends and low frequencies), we use an elliptic filter with a cut-off frequency of $0.8~$Hz and an attenuation of $40~$dB which is applied to the forward and reverse direction to avoid phase distortions. We additionally include a notch filter after observing that some ECGs are distorted by power line noise. The notch filter removes the $50~$Hz with a quality factor of $30$. Also this filter is applied to the forward and reverse direction for the same reason.
We use the pre-processing from the public library \href{https://github.com/antonior92/ecg-preprocessing}{\url{github.com/antonior92/ecg-preprocessing}}.

\begin{figure}[t]
    \centering
    \includegraphics[width=0.4\columnwidth]{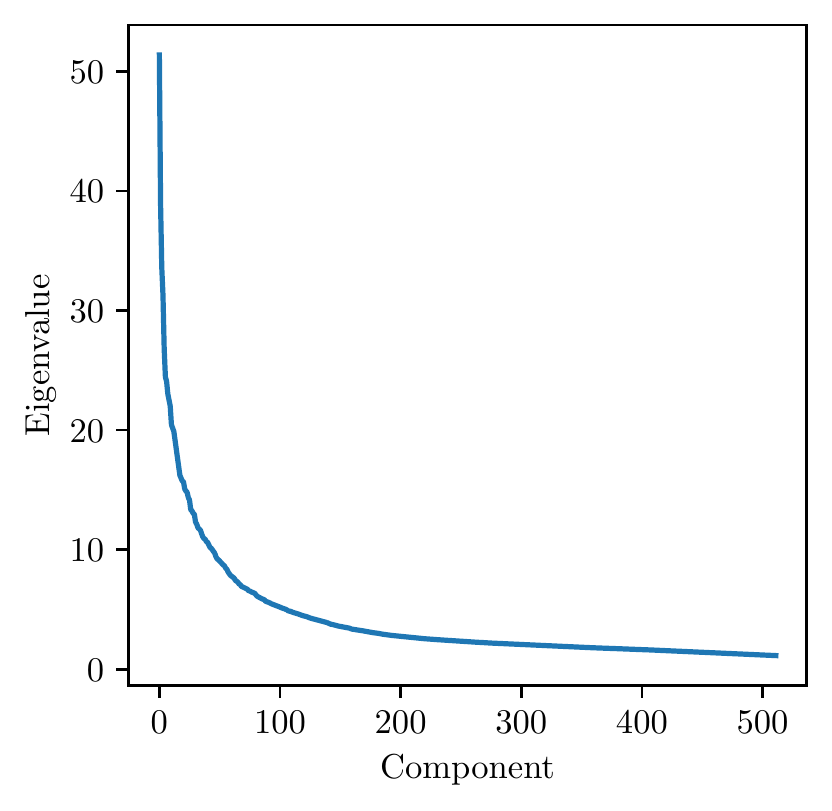}
    \caption{Eigenvalues of PCA components fit on train set. We show the first 512 eigenvalues of possible $8\cdot 4096=32\thinspace768$ ones. We choose to reduce the dimensionality of our signal to 256 as it covers most information according to this figure.}
    \label{fig:Cum_Ex_Var}
\end{figure}

For the traditional machine learning methods, to which we compare in \Cref{Results:Regression}, we further apply Prinicpal Components Analysis (PCA) to reduce the dimensionality of the data. Here, we first concatenate all leads to get a 1D signal of length $channels \times samples$. Then we fit PCA on our train dataset. We choose the number of principal components based on the eigenvalues in \cref{fig:Cum_Ex_Var}. We see that the eigenvalues decrease fast and start to converge between 200 and 300, which is why we choose to use 256 components.

\newpage
\subsection{Training Details}
    \label{appendix:TrainingDetails}
        
    \subsubsection{Network Architecture}
        We use a modified ResNet which was first developed in \cite{ribeiro2020automatic}, and later also in \cite{lima2021deep}, who also provide a public github repository: \href{https://github.com/antonior92/ecg-age-prediction}{\url{https://github.com/antonior92/ecg-age-prediction}}. We adjust the last linear layer of the model for the different task, for example different number of outputs for classification.

    \subsubsection{Hyperparameters}
        We use the the default training hyperparameters from the original network architecture repository. The only deviation is the number of epochs which we reduced from 70 to 30 since this is sufficient for our datasets to converge. The exact hyperparameters are listed in \Cref{tab:hyperparam training}.

        \begin{table}[H]
            \centering
            \caption{Hyperparameters for training the DNNs}
            \label{tab:hyperparam training}
            \begin{tabular}{l|l}
                \toprule
                Hyperparameter & Value \\
                \midrule
                optimizer & Adam\\
                maximum epochs & 30\\
                batch size & 32 \\
                initial learning rate & $10^{-3}$\\
                learning rate scheduler & ReduceLROnPlateau\\
            	patience & 7 \\
            	min. learning rate & $10^{-7}$\\
            	learning rate factor & 0.1\\
                \bottomrule
            \end{tabular}
        \end{table}

\subsection{Additional Results}
\label{appendix:additional_results}

\begin{figure}[H]
        \includegraphics{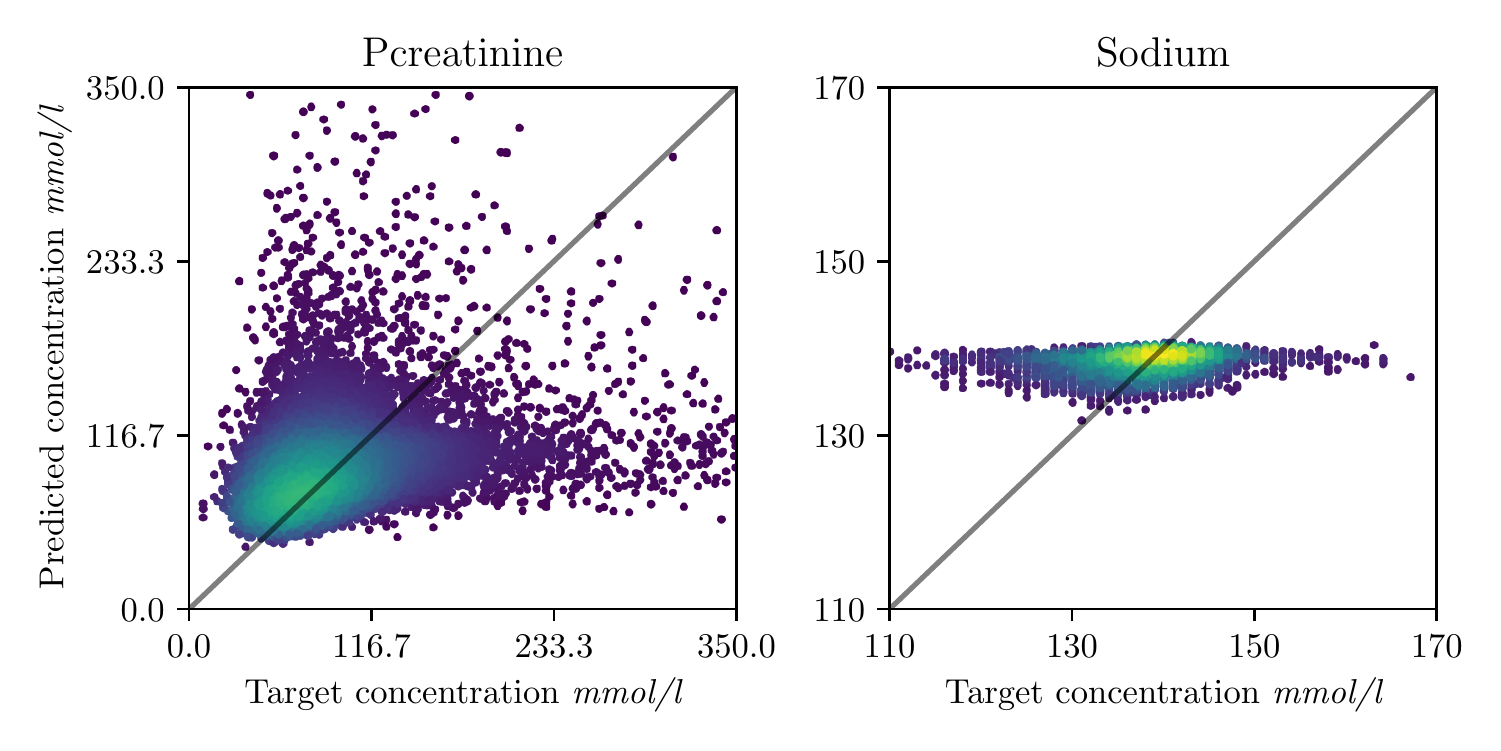}
        \centering
        \caption{\textbf{Regression scatter plot}: For sodium and creatinine, same as \cref{figure:Regression}.}
        \label{figure:AppendixRegressionSodiumPcreatintine}
    \end{figure}

%\hl{Add at least some text to these subsections, refer to all figures/tables TODO! TODO! TODO!}
Below we present additional results. First, we list more regression results with the complementing scatter plots of \cref{figure:Regression} (potassium and calcium) in \cref{figure:AppendixRegressionSodiumPcreatintine} (creatinine and sodium). Further we have a detailed performance table (more detailed than \Cref{table:Regression}) for all electrolytes in \Cref{table:AppendixRegression} for the random test set and in \Cref{table:AppendixRegressionTemporalTest} for the temporal test set. No significant difference in performance between the test sets is observed which shows that our model is robust to shift and trends over time.

\begin{figure}[H]
            \includegraphics{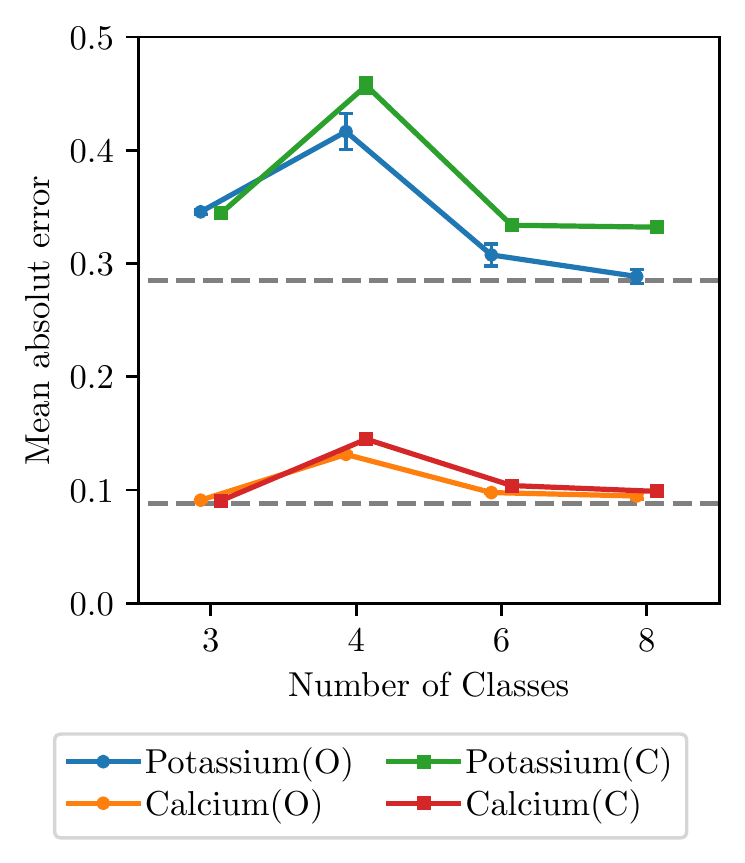}
            \centering
            \caption{\textbf{Classification (C) and Ordinal regression (O) MAE}: Similar to \cref{figure:ROC_Classification}, we plot the MAE against the number of classes. Dashed line is the MAE of the corresponding deep direct regression model.}
            \label{figure:AppendixMaePotassiumCalcium}
        \end{figure}

\begin{figure}[H]
        \includegraphics{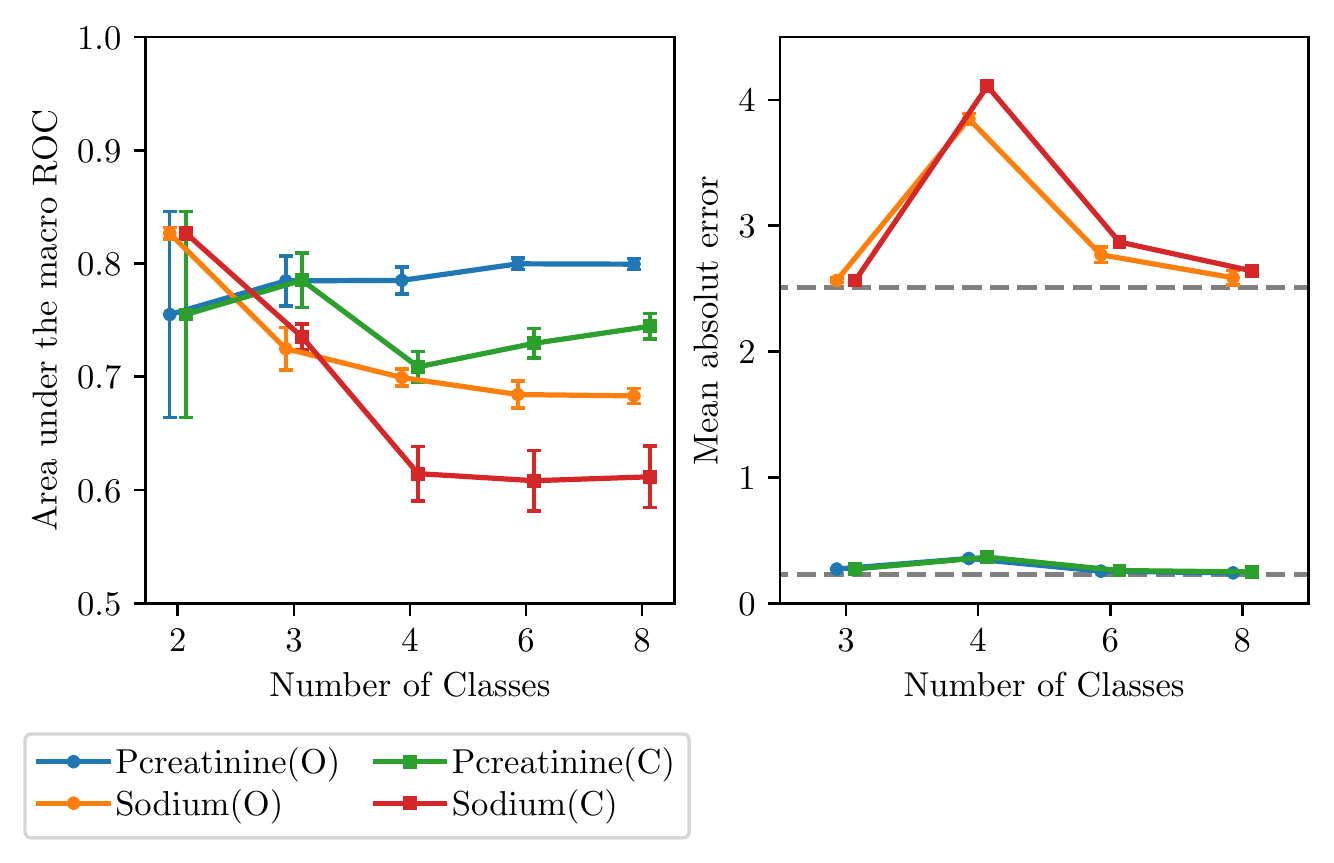}
        \centering
        \caption{\textbf{Classification (C) and Ordinal (O) regression}: Same plot as \cref{figure:ROC_Classification} and \cref{figure:AppendixMaePotassiumCalcium} but for creatinine and sodium (here we only used 4 seeds for shown mean and sd).}
        \label{figure:AppendixMaePcreatinineSodium}
    \end{figure}

Second, we list more results for classification and ordinal regression. In \cref{figure:AppendixMaePotassiumCalcium} we show the MAE for potassium and calcium which complements \cref{figure:ROC_Classification} that shows the Macro ROC. \cref{figure:AppendixMaePcreatinineSodium} complements the electrolytes by showing the Macro ROC and MAE for the other electrolytes (creatinine and sodium).

\begin{figure}[H]
            \includegraphics{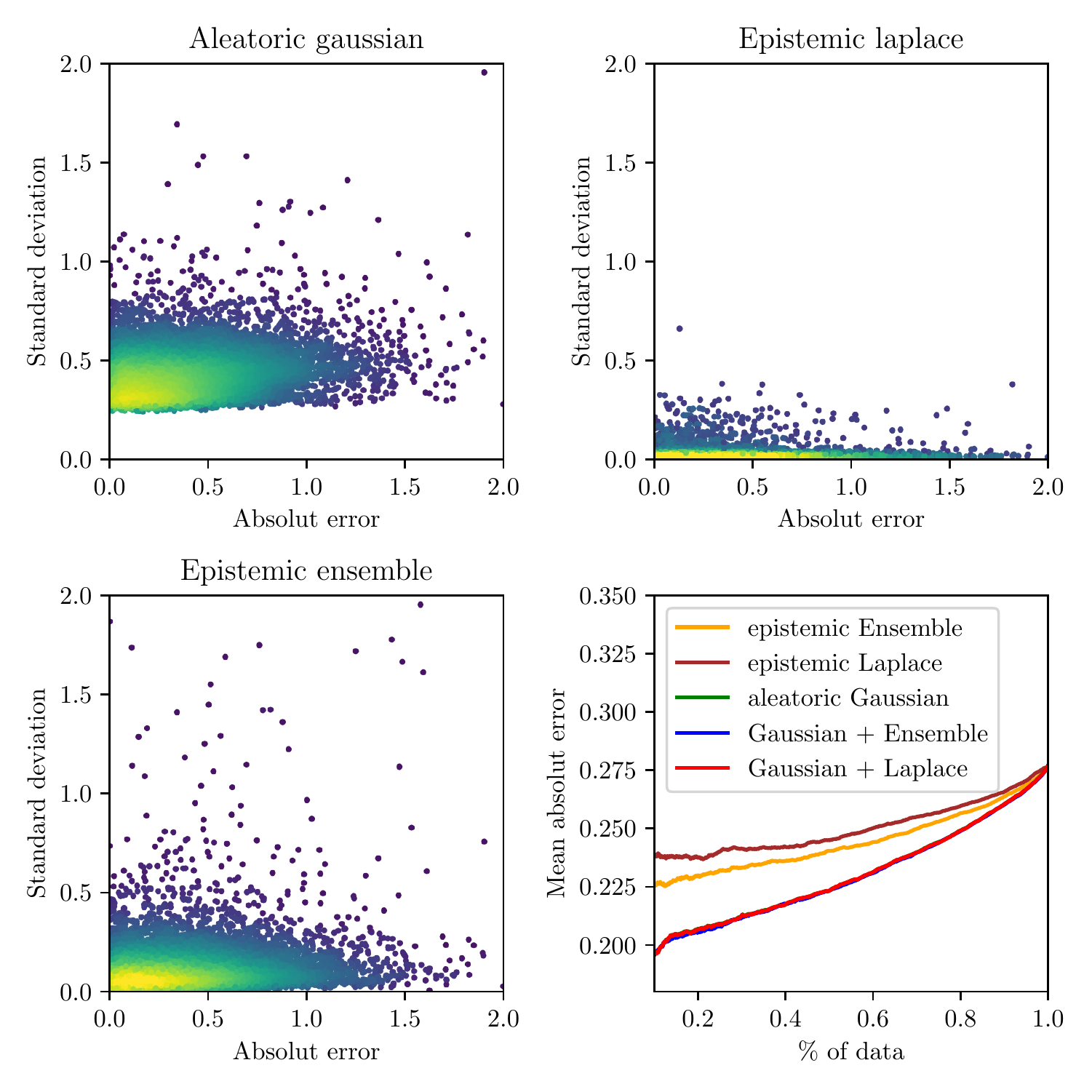}
            \centering
            \caption{\textbf{Calibration plot, potassium}: \textit{Top row and bottom left} plots show calibration plots as standard deviation vs. absolute error (to have the same units) for different uncertainties. Colors indicate frequency by a fitted Gaussian kernel density estimate. A perfectly calibrated model would follow the diagonal. \textit{Bottom right}: sparsification plot with more results than in the main paper.}
            \label{figure:AppendixCalibration}
        \end{figure}

\begin{figure}[H]
            \includegraphics{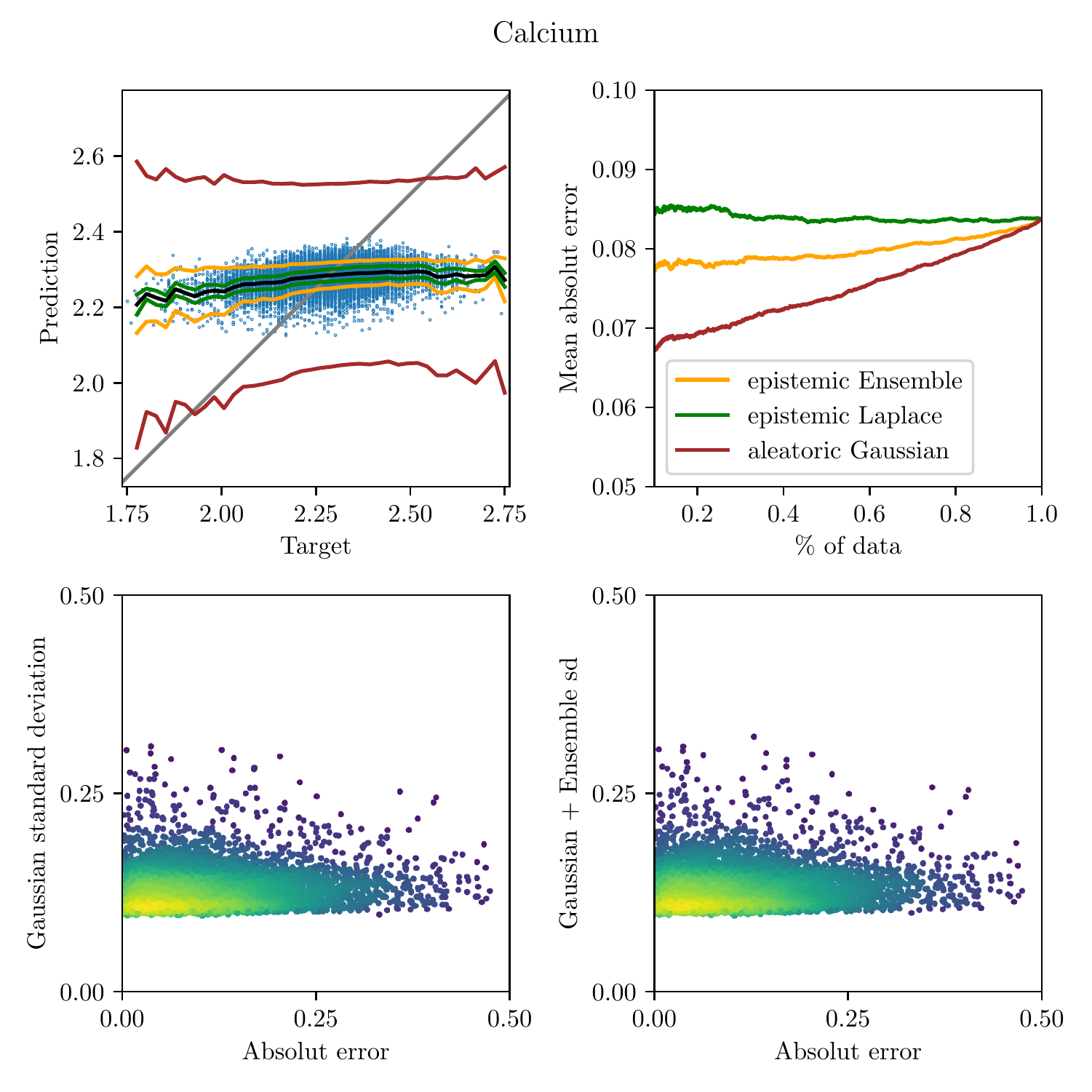}
            \centering
            \caption{\textit{Top left}: prediction vs target plot including various uncertainties. \textit{Top right}: sparsification plot. \textit{Bottom}: Calibration plots between different uncertainties and absolute error. Frequency of samples are highlighted by color which is fitted with a Gaussian kernel density estimate.}
            \label{figure:appendix:calcium prob.}
        \end{figure}

\begin{figure}[H]
            \includegraphics{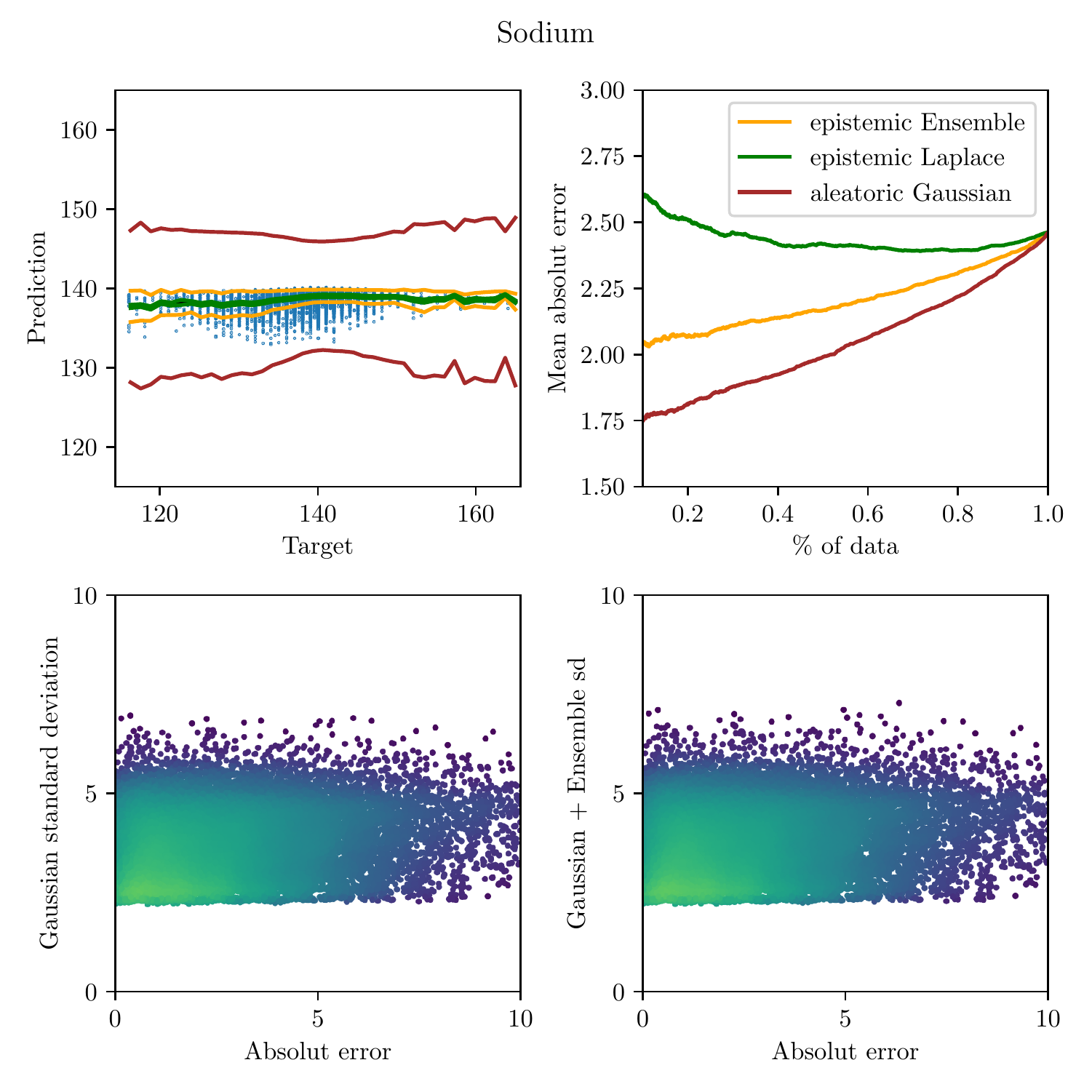}
            \centering
            \caption{Same results as \cref{figure:appendix:calcium prob.} but for sodium.}
            \label{figure:appendix:sodium prob.}
        \end{figure}

\begin{figure}[H]
        \includegraphics{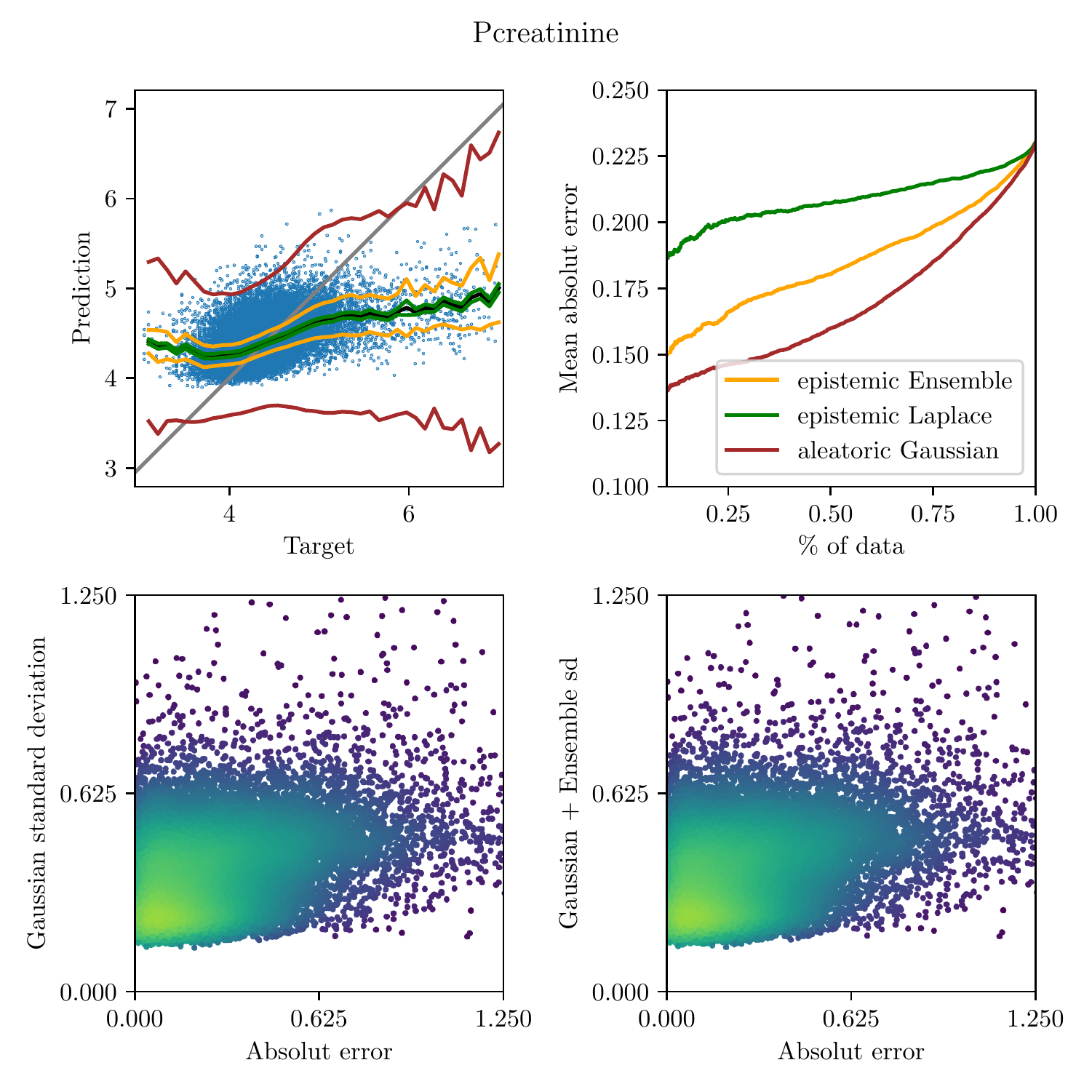}
            \centering
            \caption{Same results as \cref{figure:appendix:calcium prob.} but for creatinine, in the log transformed space due to the heavily skewed distribution of creatinine.}
            \label{figure:appendix:pcreatinine prob.}
        \end{figure}

Third, we show additional results for probabilistic regression. \cref{figure:AppendixCalibration} gives the calibration plot for potassium. The tables in \Cref{table:AppendixSparsification} and \Cref{table:AppendixCorrelation} contain numeric details about the sparsification plot for more uncertainties and the correlation between MSE and the variance to quantify the uncertainty calibration. \Cref{table:appendixOOD} lists the results of the OOD experiments. While the experiments for the SNR are expected (larger MAE and uncertainties for lower SNR), the results for masking are not as clear. While the MAE still increases, notably especially the epistemic ensemble uncertainty decreases. This means that there is less variance in the mean predictions between the different ensemble members.
Finally, \cref{figure:appendix:calcium prob.}, \cref{figure:appendix:sodium prob.} and \cref{figure:appendix:pcreatinine prob.} yield the results for the remaining electrolytes that were previously shown for potassium alone.

\newpage
%\subsubsection{Regression} 
%\label{appendix:Regression}

    \begin{table}[H]
        \centering
        \caption{\textbf{Regression performance on random test dataset}: Table shows metrics for different electrolytes of the regression models from \Cref{Results:Regression}. Target variance refers to the variance of the dataset and therefore yields a worst case MSE performance (since a model with that MSE just predicts the mean of the dataset).}
        %\resizebox{\textwidth}{!}{
        \begin{tabular}{l|cccc}%
            \toprule
            \bfseries Electrolyte & \bfseries MSE (sd) & \bfseries MAE (sd) & \bfseries Target variance & \bfseries normalized MSE (sd)% specify table head
            \begin{filecontents*}{tables/regression_table.csv}
Electrolyte,MSEMean,MSEStd,MAEMean,MAEStd,TargetVar,NormalizedMSEMean,NormalizedMSEStd,NormalizedMAEMean,NormalizedMAEStd,NormalizedTargetVar
Calcium,0.0148,0.0002,0.0877,0.0005,0.0159,0.8625,0.0088,0.6695,0.0037,0.9288
Pcreatinine,3719.8727,86.0351,26.6929,1.118,4074.4715,0.7097,0.0164,0.3687,0.0154,0.7774
Potassium,0.1524,0.0259,0.2846,0.0152,0.2158,0.6013,0.1021,0.5653,0.0301,0.8517
Sodium,12.5933,0.1108,2.5123,0.0156,13.1026,0.8445,0.0074,0.6506,0.004,0.8787
\end{filecontents*}
\csvreader[head to column names]{tables/regression_table.csv}{}% use head of csv as column names
            {\\\midrule\Electrolyte & \MSEMean(\MSEStd) & \MAEMean (\MAEStd) & \TargetVar &\NormalizedMSEMean(\NormalizedMSEStd)}\\% specify your columns here
            \bottomrule
        \end{tabular}
        %}
        \label{table:AppendixRegression}
    \end{table}
    
    \begin{table}[H]
        \centering
        \caption{\textbf{Regression performance on  temporal test dataset}: Table shows metrics for different electrolytes of the regression models from \Cref{Results:Regression}. Target variance has same meaning as in \Cref{table:AppendixRegression}.}
        % \resizebox{\textwidth}{!}{
        \begin{tabular}{l|cccc}%
            \toprule
            \bfseries Electrolyte & \bfseries MSE (sd) & \bfseries MAE (sd) & \bfseries Target variance & \bfseries normalized MSE (sd)% specify table head
            \begin{filecontents*}{tables/regression_table_temporal_test.csv}
Electrolyte,MSEMean,MSEStd,MAEMean,MAEStd,TargetVar,NormalizedMSEMean,NormalizedMSEStd,NormalizedMAEMean,NormalizedMAEStd,NormalizedTargetVar
Calcium,0.0201,$\leq$1e-4,0.1059,0.0003,0.0167,1.1742,0.0028,0.8086,0.0019,0.973
Pcreatinine,2973.3104,156.24,24.5017,1.2979,3370.132,0.5673,0.0298,0.3384,0.0179,0.643
Potassium,0.1319,0.0171,0.262,0.0127,0.1987,0.5203,0.0675,0.5204,0.0252,0.7842
Sodium,12.0118,0.084,2.3903,0.0093,12.7672,0.8055,0.0056,0.619,0.0024,0.8562
\end{filecontents*}
\csvreader[head to column names]{tables/regression_table_temporal_test.csv}{}% use head of csv as column names
            {\\\midrule\Electrolyte & \MSEMean(\MSEStd) & \MAEMean (\MAEStd) & \TargetVar &\NormalizedMSEMean(\NormalizedMSEStd)}\\% specify your columns here
            \bottomrule
        \end{tabular}
        % }
        \label{table:AppendixRegressionTemporalTest}
    \end{table}

\newpage
%\subsection{Classification and Ordinal Regression}
%\label{appendix:Classification}

%\subsection{Probabilistic Regression}
%\label{appendix:Probabilistic Regression}
\newpage
    %\subsubsection{Potassium}
        
        % \begin{table}[H]
        %     \centering
        %     \caption{\textbf{Sparsification against MAE}: Numbers in columns show different levels of sparsification (in percent), corresponding row values MAE. This table gives the numeric values of bottom right plot of \cref{figure:AppendixCalibration}.}
        %     \resizebox{\textwidth}{!}{\begin{tabular}{l|cccc}% wrap row, we can also change labels again if preferred
        %         \toprule
        %         \bfseries  & \bfseries 25 & \bfseries 50 & \bfseries 75 & \bfseries 100  % specify table head
        %         \csvreader[head to column names]{tables/Sparsification_Correlation_table.csv}{}% use head of csv as column names
        %         {\\\midrule \Electrolyte & \SparTwentyfiveMean(\SparTwentyfiveStd) & \SparFiftyMean (\SparFiftyStd) & \SparSeventyfiveMean(\SparSeventyfiveStd) & \SparOnehundredMean(\SparOnehundredStd)}\\% specify your columns here
        %         \bottomrule
        %     \end{tabular}}
        %     \label{table:AppendixSparsification}
        % \end{table}
        
        \newpage
        \begin{table}[H]
            \centering
            \caption{\textbf{Sparsification against MAE}: Numbers in columns show different levels of sparsification (in percent), the corresponding row shows MAE values. This table gives the numeric values of the bottom right plot of \cref{figure:AppendixCalibration}.}
            % \resizebox{\textwidth}{!}{
            \begin{tabular}{l|llll}% wrap row, we can also change labels again if preferred
                \toprule
                 & \textbf{25} & \textbf{50} & \textbf{75} & \textbf{100} \\
                \midrule
                Aleatoric Gaussian & 0.213(0.007) & 0.228(0.008) & 0.250(0.009) & 0.283(0.009) \\
                Epistemic ensemble & 0.235(0.006) & 0.246(0.009) & 0.259(0.010) & 0.283(0.009) \\
                Epistemic Laplace  & 0.249(0.014) & 0.260(0.021) & 0.271(0.019) & 0.283(0.009) \\
                \midrule
                Aleatoric Gaussian + Epistemic ensemble & 0.211(0.005) & 0.227(0.007) & 0.249(0.008) & 0.283(0.009) \\
                Aleatoric Gaussian + Epistemic Laplace & 0.212(0.007) & 0.228(0.008) & 0.250(0.009) & 0.283(0.009) \\
                \midrule
                Epistemic ensemble of direct reg. & 0.227(NA) & 0.238(NA) & 0.249(NA) & 0.274(NA) \\
                \bottomrule
            \end{tabular}
            % }
            \label{table:AppendixSparsification}
        \end{table}
        
        % \begin{table}[H]
        %     \centering
        %     \caption{\textbf{Correlation between MSE and Variance}: we correlate the MSE with the variance from different uncertainties. A correlation of 1 would yield perfect calibration.}
        %     \begin{tabular}{l|l}%
        %         \toprule
        %         \bfseries  & \bfseries Correlation% specify table head
        %         \csvreader[head to column names]{tables/Sparsification_Correlation_table.csv}{}% use head of csv as column names
        %         {\\\midrule \Electrolyte &\CorrelationMean(\CorrelationStd)}\\% specify your columns here
        %         \bottomrule
        %     \end{tabular}
        %     \label{table:AppendixCorrelation}
        % \end{table}
        
        \begin{table}[H]
            \centering
            \caption{\textbf{Correlation between MSE and Variance}: we correlate the MSE with the variance from different uncertainties. A correlation of 1 would indicate perfect calibration.}
            \begin{tabular}{l|l}%
                \toprule
                 & Correlation \\
                \midrule
                Aleatoric Gaussian & 0.225(0.066) \\
                Epistemic ensemble & 0.218(0.010) \\
                Epistemic Laplace  & 0.068(0.034) \\
                \midrule
                Aleatoric Gaussian + Epistemic ensemble & 0.255(0.039) \\
                Aleatoric Gaussian + Epistemic Laplace & 0.225(0.066) \\
                \midrule
                Epistemic ensemble of direct reg. & 0.225(NA) \\
                \bottomrule
            \end{tabular}
            \label{table:AppendixCorrelation}
        \end{table}

    \begin{table}[H]
        \centering
        \caption{\textbf{OOD experiments}: This is an extended table from \Cref{table:OOD}. SNR X refers to OOD experiments with varying SNR; Mask X refers to OOD experiments where X percent of the data is masked.}\label{table:appendixOOD}
        {\begin{tabular}{l|l|llll}%
            \toprule
             & MAE & \shortstack{Aleatoric\\Gaussian} & \shortstack{Epistemic\\ensemble} & \shortstack{Epistemic\\Laplace} & \shortstack{Epistemic\\direct reg.}\\
            \midrule
            Baseline & 0.304(0.021) & 0.389(0.012) & 0.121(0.048) & 0.022(0.003) & 0.099 \\
            \midrule
            SNR 10 & 0.330(0.016) & 0.399(0.012) & 0.149(0.041) & 0.028(0.009) & 0.134 \\
            SNR 1  & 0.368(0.026) & 0.480(0.078) & 0.184(0.075) & 0.049(0.031) & 0.154 \\
            \midrule 
            Mask 25 & 0.300(0.008) & 0.386(0.015) & 0.091(0.015) & 0.022(0.001) & 0.098 \\
            Mask 50 & 0.311(0.005) & 0.388(0.008) & 0.070(0.010) & 0.020(0.002) & 0.073 \\
            Mask 75 & 0.334(0.001) & 0.385(0.004) & 0.047(0.005) & 0.018(0.002) & 0.054 \\
            \bottomrule
        \end{tabular}}
    \end{table}

\end{document}